\documentclass[twocolumn,prd,aps,tightenlines,floats,floatfix,preprintnumbers,nofootinbib,eqsecnum]{revtex4}

\def\sp{\kern +3pt}
\def\sm{\kern -3pt}
\def\spQ{\kern +6pt}

\def\bea{\begin{eqnarray}}
\def\eea{\end{eqnarray}}

\def\sfrac#1#2{{\textstyle \frac{#1}{#2}}}

\newcommand{\ket}[1]{|#1\rangle}

\def\be{\begin{equation}}
\def\ee{\end{equation}}
\def\ba{\begin{eqnarray}}
\def\ea{\end{eqnarray}}

\usepackage{graphics}
\usepackage{graphicx}
\usepackage{epsf}
\usepackage{amsmath}
\usepackage{amssymb}
\usepackage{bbold}

\usepackage{xcolor}

\usepackage{fnbreak}

\begin{document}

\phantom{0}
\vspace{-0.2in}
\hspace{5.5in}

\preprint{{\bf  LFTC-20-1/53}}

\vspace{-1in}

\title
{\bf 
Covariant model for Dalitz decays of decuplet baryons to octet baryons}
\author{G.~Ramalho}
\vspace{-0.1in}

\affiliation{Laborat\'orio de 
F\'{i}sica Te\'orica e Computacional -- LFTC,
Universidade Cruzeiro do Sul and Universidade Cidade de  S\~ao Paulo,  \\
01506-000,   S\~ao Paulo, SP, Brazil}

\vspace{0.2in}
\date{\today}

\phantom{0}

\begin{abstract}
In the last years it became possible to 
measure in HADES the dilepton decays of several baryons.
The baryon dilepton decays provide information about 
the electromagnetic structure of the baryons in the timelike region.
In the present work, we study the $B^\prime \to e^+ e^- B$ decays, 
where $B^\prime$ is a baryon decuplet member and $B$ is a baryon octet member.
Our calculations are based on the covariant spectator 
quark model, where the contribution of the quark core 
is complemented with an $SU(3)$ contribution from the pion cloud.
The pion cloud contribution prove 
to be relevant in the range of study.
We present predictions 
for the $\Sigma^0 (1385) \to e^+ e^- \Lambda(1116)$
and $\Sigma^{+}(1385) \to e^+ e^- \Sigma^+(1193)$ decays,
which may be tested at HADES in a near future.
Predictions for the remaining decuplet baryon Dalitz decays 
are also presented.
We conclude that different orders of magnitudes 
are expected for the baryon decuplet Dalitz decay widths, 
according to the quark content of the baryons.
We also conclude that the dependence of 
the transition form factors 
on the square momentum transfer ($q^2$) 
is important for some transitions.
\end{abstract}

\vspace*{0.9in}  
\maketitle

\section{Introduction}
\label{secIntro}

In the last decades, there was a significant progress 
in the study of the electromagnetic structure 
of the nucleon ($N$)
and nucleon excitations ($N^\ast$)~\cite{NSTAR,Aznauryan12,NSTAR17}.
Most of the measured data were obtained 
through the scattering of electrons on nucleon targets 
($e^- N \to e^- N^\ast$ ),
which probes the region where the square
four-momentum transfer $q^2$ is negative 
($q^2 < 0$), also known as the spacelike region.
In the electron scattering experiments 
the analysis of the data is based on 
the $\gamma^\ast N \to N^\ast$ transition,
where the spacelike virtual photon is produced 
by the incoming electron,
and the $\gamma^\ast N \to N^\ast$ transition form factors
are extracted from the experimental cross sections.
Experiments based on electron-nucleon scattering 
have been performed in facilities such as 
Jefferson Lab, MIT-Bates, ELSA, MAMI among others,
to probe the electromagnetic structure 
of $N^\ast$ states in the 
first three resonance regions~\cite{NSTAR,Aznauryan12,Drechsel07,Burkert04,NSTAR17,PDG18}.

The electron scattering technique can also
be used to probe the electromagnetic 
structure of the hyperons (baryons with strange quarks)
based on the $\gamma^\ast B \to B^\prime$ transitions, 
where $B$ and $B^\prime$ are generic hyperons.
In practice, however, the technique is almost exclusively limited
to  nucleon targets, since hyperons targets 
are difficult to produce due to their short lifetime,
except in the limit $q^2 = 0$.
In that limit, there are measurements 
of magnetic moments of a few hyperons and some magnetic transition
moments~\cite{Hyperons,Octet4,Octet1,Octet2,LambdaSigma0,Octet2Decuplet,DecupletDecays,PDG18}.  
Another limitation of the electron scattering technique
is that it is restricted to the $q^2 \le 0$ region.

The timelike region ($q^2 > 0$) can be accessed at HADES (GSI)
through some exclusive reaction channels in proton-proton ($pp$)
collisions or by pion-induced reactions~\cite{HADES14,HADES17,HADES17b,ColeTL,Salabura13,Weil12,Ramstein18a,Ramstein18b,HADES17c,Ramstein19,Lutz03,Shyam10,Zetenyi12,HADES20c,Salabura20}.
In the proton-proton collisions 
the channel $pp \to p p  e^+ e^-$ 
probes the structure of the intermediate $N^\ast$ states
through the elementary reactions
$N^\ast \to p \gamma^\ast \to p  e^+ e^-$~\cite{HADES14,HADES17,Dohrmann10}.
The $\Delta(1232)$ Dalitz decay was recently analyzed at HADES based on the study of 
the $pp \to pp  e^+ e^-$ channel on $pp$ scattering~\cite{HADES17,HADES17b}.
The results were compared our estimates~\cite{Timelike2}.
The pion-induced reactions, are particularly important to 
study $N^\star$ resonances which decay 
into two or more pions~\cite{Ramstein18a,Ramstein18b,HADES17c,Ramstein19,HADES20c}.
Measurements of the $N(1520)$ and $N(1535)$ Dalitz decays are  
in progress at HADES~\cite{N1520TL,N1535TL,Ramstein18a,Ramstein18b,Przygoda16,Scozzi17}.
In both methods, we access the region  $4 m_e^2 \le  q^2 \le (M_{B'} -M_N)^2$,
where $m_e$ is the electron mass, and $M_{B'}$ and $M_N$ are the $N^\ast$ and nucleon masses, 
respectively.
The production of timelike photons is   
clearly identified by the detection of
 $e^+ e^-$ pairs (dileptons)\footnote{Although 
the term dilepton can be used for muon pairs ($\mu^+ \mu^-$), 
we follow the usual nomenclature and use 
dilepton to refer to a electron-positron pair ($e^+ e^-$).}
in the final state,
due to the conversion $\gamma^\ast \to e^+ e^-$.
Experiments at HADES complement then the experiments based 
on electron-nucleon scattering, in the spacelike region ($q^2 \le 0$)~\cite{ColeTL,Briscoe15}.

Another timelike sub-region, not discussed in the present work,
is the region probed by $e^+ e^-$ and $p \bar p$ collisions 
at BaBar, BES-III, CLEO, and 
PANDA/FAIR~\cite{Pacetti15a,Aubert07a,Dobbs17a,Ablikim18a,Singh16a},
which access the baryon ($B$) elastic form factors when $q^2 \ge 4 M_B^2$, 
where $M_B$ is the baryon mass~\cite{Hyperons,Omega3}.

HADES provides a unique opportunity to explore 
the electromagnetic structure of baryons
based on the $B' \to \gamma^\ast B$ transitions,
where $B'$ and $B$ are generic baryons, through the dilepton decays 
($B' \to e^+ e^- B$)~\cite{HADES14,HADES17,HADES17c,Ramstein18a,Ramstein18b,ColeTL,Ramstein19}.
Different from the traditional electron-nucleon scattering, 
at HADES one can probe the electromagnetic structure of the
hyperons in the kinematic region $4 m_e^2 \le q^2 \le (M_{B'} -M_B)^2$,
where $M_{B'}$, $M_B$ are the baryon masses~\cite{Briscoe15,Lalik19,Ramstein19}.
Measurements of strangeness production are possible 
due to the large acceptance and excellent particle identification, 
including dileptons in the final state~\cite{Salabura13}.
In progress are feasibility studies on the $\Sigma (1385)$, $\Lambda (1404)$ 
and $\Lambda (1520)$ Dalitz decays
by the HADES collaboration~\cite{Lalik19,Rathod20,HADES20}.
Those studies suggest that those decays 
can be measured at GSI in the next few years 
and subsequently also at FAIR~\cite{Lalik19,Ramstein18a,ColeTL,Junker20a,Holmberg18}.

From the theoretical side there 
are not many models available
for baryon electromagnetic 
transitions in the timelike region~\cite{Weil12,Kaxiras85,Williams93,Sakurai60,Krivoruchenko02,Iachello04,Lutz03,Zetenyi12,Junker20a,Shyam10}.
An important constraint on those models 
is that the transition between the spacelike region 
and the timelike region (interval between $q^2=0$ and $q^2 = 4 m_e^2$)
must be smooth~\cite{Ramstein18b,ColeTL,Briscoe15}.
There are a few theoretical issues, 
which need to be discussed:
What happens in the transition between the spacelike region ($q^2 \le 0$),
and the timelike region ($q^2 > 0$)
where relevant imaginary components 
appear on the transition form factors  
above the two-pion threshold  ($q^2 > 4 m_\pi^2$) 
for isovector transitions, 
and above the three-pion threshold ($q^2 > 9 m_\pi^2$) 
for isoscalar transitions ($m_\pi$ is the pion mass).
How important are the physical poles 
associated with the meson resonances.
How significant is the $q^2$ dependence of the form factors, 
and how are form factors modified 
near the pseudothreshold 
$q^2= (M_{B'} -M_B)^2$~\cite{HADES17,Dohrmann10,Siegert4,Siegert1,Siegert2,GlobalFit}.

In the spacelike region, including the limit $Q^2=0$,
there are calculations based on 
nonrelativistic and relativistic quark models~\cite{Kaxiras85,Darewych83,Sahoo95,Wagner98,Bijker00,Keller12},
Dyson-Schwinger equations~\cite{Alepuz16a,Alepuz18},
lattice QCD simulations~\cite{Leinweber93},  
QCD sum rules~\cite{Aliev06,Wang09}, 
Skyrme and soliton models~\cite{Schat95,Haberichter97,Kim20a}, 
chiral perturbation theory
and large $N_c$ limit~\cite{Butler93a,Arndt04,Holmberg18,Lebed11}.

From the analysis of the spacelike data, 
one can conclude that models based strictly on the quark degrees of freedom 
are insufficient to explain the measured transition form factors.
The effects associated with the meson cloud dressing 
of the bare cores are crucial to describe 
the data in the region $0 \ge  q^2 > - 2$ GeV$^2$, 
as demonstrated already for the 
$\Delta(1332)$~\cite{NSTAR,Aznauryan12,NSTAR17,Burkert04,NDelta,NDeltaD,LatticeD}.
Our model for the  $\Delta(1232)$ Dalitz decay~\cite{Timelike2},
which describe the HADES data~\cite{HADES17},
corroborates also the importance of the of pion cloud for the 
$\gamma^\ast N \to \Delta(1232)$ transition
in the timelike region, as in the spacelike region.
There is therefore a great interest in studying 
the roles of the valence quark and meson cloud effects 
in the timelike region~\cite{Timelike,Timelike2,NSTAR17,N1520TL,N1535TL}.

Motivated by the experiments planned for HADES,
in the present work we focus on the $B' \to \gamma^\ast B$ transitions, 
where $B^\prime$ is a baryon decuplet member and $B$ is a baryon octet member
(decuplet baryon decays).
We restrict for now our study to baryon systems 
that best fit an $SU(3)$ quark model classification
(baryon octet and baryon decuplet). 
Our calculations are based on the covariant 
spectator quark model~\cite{NSTAR17,Nucleon,Omega} 
developed previously for the $\gamma^\ast B \to B^\prime$ transitions in 
the spacelike region~\cite{DecupletDecays}.

The covariant spectator quark model provides 
an alternative to valence quark models 
which do not take into account meson cloud excitations 
of the bare cores, and simplified vector meson dominance (VMD) 
models~\cite{Weil12,Sakurai60,Krivoruchenko02,Iachello04},
which do not take into account the underlying quark substructure 
of the baryons.
The formalism has been used in the study 
of the electromagnetic and the axial structure of 
the nucleon, several nucleon excitations, 
and hyperons~\cite{Nucleon2,NDelta,NDeltaD,Delta1600,Lattice,LatticeD,LambdaSigma0,Octet1,Octet2,Jido12,DeltaFF,Roper,SRapp,Hyperons,Nexcitations,OctetAxial,N1535TL,N1520TL,Timelike,Timelike2,Nucleon,Omega,Octet4}.

The covariant spectator quark model of the $\gamma^\ast B \to B^\prime$ transition~\cite{DecupletDecays}
is extended in the present work to the timelike region.
Within the formalism, the octet baryon to decuplet baryon 
electromagnetic transitions 
are dominated by the
magnetic transition form factor~\cite{NSTAR17,Nucleon,Omega},
which can be decomposed into valence quark  
and meson cloud contributions~\cite{DecupletDecays,Octet2Decuplet}.
The meson cloud contribution is calculated 
from a microscopic pion-baryon model, calibrated by 
the $\gamma^\ast N \to \Delta(1232)$ 
transition, and extended to the octet baryon to decuplet baryon
electromagnetic transitions~\cite{Timelike2,Octet2Decuplet,DecupletDecays}.

We use our formalism to estimate the baryon decuplet 
Dalitz decay widths in terms of the 
square invariant mass of the dilepton pair $q^2$,  
and the square invariant energy $W^2$ of the $\gamma^\ast B$ system~\cite{HADES14,Dohrmann10}.
We present, in particular, predictions for the 
$\Sigma^0 (1385) \to e^+ e^- \Lambda(1116)$ and 
$\Sigma^+ (1385) \to e^+ e^- \Sigma^+ (1193)$ decays, 
which may be tested by future HADES experiments~\cite{Lalik19,ColeTL}.
As for the remaining decays, 
we estimate that the magnitudes of the 
$\Xi^0 (1530) \to e^+ e^- \Xi^0 (1318)$ and 
$\Sigma^+ (1385) \to e^+ e^- \Sigma^+ (1193)$ decay widths 
are comparable to the magnitude of the $\Delta(1232) \to e^+ e^- N$ decay width,
as suggested by $SU(3)$ and $U$-spin estimates~\cite{DecupletDecays,Keller12}.
We present also calculations for the radiative decay widths 
in terms of the invariant mass $W$, 
and compare our estimates with the available data.
We conclude also that 
our estimate of the $\Sigma^- (1385) \to \gamma \, \Sigma^- (1193)$ width,
unknown at the moment,
is close to the present experimental limit,
and may therefore be measured in a near future.

This article is organized as follows:
In the next section, we review the formalism 
associated with the radiative and Dalitz decays 
of $3/2^+$ baryons into $1/2^+$ baryons.
The covariant spectator quark model is discussed in Sec.~\ref{secFormalism},
where we present also numerical results for the transition form factors.
Our results for the radiative and Dalitz decays of 
the decuplet baryons $B^\prime$ are presented and 
discussed in Sec.~\ref{secDalitz}.
The outlook and conclusions are presented in Sec.~\ref{secConclusions}.


\section{Dalitz decay of decuplet baryons}

A baryon $B^\prime$ can decay in different channels,
including meson-baryon states, (multi-meson)-baryon states,
the radiative decay ($\gamma \, B$) and the dilepton decay ($e^+ e^- B$).
In the present section, we focus on the radiative ($B^\prime \to \gamma \, B$)
and dilepton ($B^\prime \to e^+ e^- B$) decays.
The formalism described below is a generalization
of the formalism for the $\Delta(1232) \to \gamma\,  N$ 
and  $\Delta(1232) \to e^+ e^- N$ 
decays~\cite{Krivoruchenko01,Dohrmann10,Timelike2,Wolf90}.

We assume that $B'$ is a member of the baryon decuplet 
(state $\frac{3}{2}^+$) 
and that $B$ a is a member of the baryon octet (state $\frac{1}{2}^+$).
Both baryons have positive parity.
As before, $M_{B^\prime}$ and $M_{B}$ represent 
the mass of $B'$ and $B$, respectively.

The Dalitz decay of the baryon  $B^\prime$ is determined 
by the function $\Gamma_{\gamma^\ast B} (q,W)$, 
where $W$ is the energy of the resonance $B^\prime$, $q=\sqrt{q^2}$ 
and $q^2$ is the virtual photon ($\gamma^\ast$) square four-momentum.
The baryon $B^\prime$ Dalitz decay is the consequence of the decay 
of the timelike virtual  photon into 
a pair of electrons ($\gamma^\ast \to e^+ e^-$).

The function  $\Gamma_{\gamma^\ast B} (q,W)$ 
can be written~\cite{Krivoruchenko01,Dohrmann10,Timelike,Timelike2} as
\ba
\Gamma_{\gamma^\ast B}(q,W) =
\frac{\alpha}{16}
\frac{(W+ M_B)^2}{W^3 M_B^2} \sqrt{y_+ y_-} y_- |G_T (q^2,W)|^2,
\nonumber \\
\label{eqGammaG}
\ea
where $\alpha \simeq 1/137$ is the fine-structure constant,
and
\ba
y_\pm = (W\pm M_B)^2 - q^2.
\ea
The function $|G_T (q^2,W)|$ depends on 
the Jones-Scadron form factors:
$G_M$ (magnetic dipole), $G_E$ (electric quadrupole)
and $G_C$ (Coulomb quadrupole)~\cite{Jones73,Devenish76},
and takes the form 
\ba
& &
|G_T (q^2,W)|^2 = \nonumber \\
& &
|G_M (q^2,W)|^2 + 3 |G_E (q^2,W)|^2 + \frac{q^2}{2 W} |G_C (q^2,W)|^2.
\nonumber \\
\label{eqGT}
\ea

The functions $\Gamma_{\gamma B} (W)$ and 
$\Gamma_{e^+ e^- B} (W)$ which quantify 
the radiative and Dalitz decays, respectively,
are calculated with the assistance
of the function  $\Gamma_{\gamma^\ast B} (q,W)$,
as discussed below.

The photon decay width is defined by 
the limit \mbox{$q^2=0$}~\cite{Dohrmann10,Wolf90}
\ba
\Gamma_{\gamma B} (W) = \Gamma_{\gamma^\ast B}(0,W).
\label{eqGamma0}
\ea

The Dalitz decay width  $\Gamma_{e^+ e^- B} (W)$ is determined by integrating 
\ba
\Gamma^\prime_{e^+ e^- B} (q,W) 
\equiv 
\frac{d \Gamma_{e^+ e^- B}}{d q} (q,W),
\ea
according to 
\ba
\Gamma_{e^+ e^- B}(W) = \int_{2 m_e}^{W -M_B} 
\Gamma^\prime_{e^+ e^- B} (q,W) dq.
\label{eqGammaInt}
\ea
In the previous equation 
the interval of integration $4 m_e^2 \le q^2 \le (W- M_B)^2$
is the consequence of the threshold 
of the dilepton production and the 
maximum value of the photon square four-momentum allowed 
by the $B^\prime \to \gamma^\ast B$ decay: $q^2= (W -M_B)^2$.
This is the value of $q^2$ obtained 
when the photon three-momentum vanishes $|{\bf q}|=0$~\cite{Timelike,Siegert1,Siegert2,Siegert4}.
The function  $\Gamma^\prime_{e^+ e^- B} (q,W) $ 
can be evaluated using~\cite{HADES17,Dohrmann10,Timelike2,Wolf90}  
\ba
\Gamma^\prime_{e^+ e^- B} (q,W) = \frac{2 \alpha}{3 \pi q}
\Gamma_{\gamma^\ast B} (q,W). 
\label{eqGammaP}
\ea

The relations (\ref{eqGamma0}), (\ref{eqGammaInt}) 
and (\ref{eqGammaP}) demonstrate that the decay widths 
$\Gamma_{\gamma B} (W)$ and $\Gamma_{e^+ e^- B} (W)$
are determined, once one has a model for 
the effective form factor $|G_T (q^2,W)|$.
Note, however, that the model should be defined 
for arbitrary values of $W$ (invariant energy of the $\gamma^\ast B$ system),
since the measurements are performed for values of 
$W$ which may differ from the decuplet baryon mass ($M_{B'}$).
Our model for  $|G_T (q^2,W)|$ is described 
in the next section.

The baryon $B'$ radiative decay ($B' \to \gamma \, B$)
measured in the experiments, correspond to
the result from Eq.~(\ref{eqGammaG})
in the limits $W= M_{B'}$ and $q^2=0$:
\ba
\Gamma_{\gamma B} \equiv 
\Gamma_{\gamma^\ast B} (0, M_{B'}).
\label{eqGammaPole}
\ea

\section{Covariant spectator quark model}
\label{secFormalism}

In the present section, we describe the formalism associated 
with the covariant spectator quark model~\cite{NSTAR17,Nucleon,Omega}.
The covariant spectator quark model was derived from 
the covariant spectator theory~\cite{Gross,Nucleon}.
In this framework a baryon is described as a three-constituent 
quark system, where a quark is free to interact 
with the electromagnetic fields.
Integrating over the internal degrees of freedom 
of the non-interacting quark-pair, one reduces 
the three-quark system to a quark-diquark system
where the spectator quark-pair is represented 
by an on-mass-shell diquark with 
an effective mass $m_D$~\cite{Nucleon,Nucleon2,Omega}. 
The effective quark-diquark wave function 
is free of singularities and include the quark confinement 
implicitly~\cite{Nucleon,NSTAR,NSTAR17,Gross}.
The wave functions of the baryons are built 
according to the spin-flavor-radial symmetries
with the radial wave functions determined phenomenologically 
by the experimental data, or lattice QCD data 
for some ground state systems~\cite{Nucleon,Omega,LatticeD,NSTAR17}.

In the electromagnetic interaction with the quarks, 
we take into account the structure 
associated with gluon and quark-antiquark dressing of the quarks.
This structure is parametrized in terms of constituent 
quark electromagnetic form factors~\cite{Nucleon,Omega}.

The covariant covariant spectator quark model
was already applied to the study of 
the electromagnetic structure of several baryons 
in the spacelike region~\cite{ Nucleon2,NDelta,NDeltaD,Delta1600,LambdaSigma0,Octet1,Jido12,DeltaFF,Roper,SRapp,Nexcitations,OctetAxial},
in the timelike region~\cite{Hyperons,Timelike,Timelike2,N1535TL,N1520TL},
to the structure of baryons in the nuclear medium~\cite{Octet2},
and to the lattice QCD regime~\cite{Lattice,LatticeD,Omega}.
We discuss next the formalism 
associated with the octet and decuplet baryons.

\begin{table*}[t]
\begin{center}
\begin{tabular}{l c c c}
\hline
\hline
$B$   & $\ket{M_A}$  & &  $\ket{M_S}$  \\ 
\hline
$p$     &    $\sfrac{1}{\sqrt{2}} (ud -du) u$
 & &
 $\sfrac{1}{\sqrt{6}} \left[
        (ud + du) u - 2 uu d \right]$  \\
$n$     &  $\sfrac{1}{\sqrt{2}} (ud -du) d$ & &
 $-\sfrac{1}{\sqrt{6}} \left[
         (ud + du) d - 2 ddu \right]$  \\[.3cm]
$\Lambda^0$ &  $\sfrac{1}{\sqrt{12}}
\left[
s (du-ud) - (dsu-usd) -2(du-du)s
\right]$
& &
$\sfrac{1}{2}
\left[ (dsu-usd) + s (du-ud)
\right]$ \\[.3cm]
$\Sigma^+$  & 
$\sfrac{1}{\sqrt{2}} (us -su) u $ 
& &  
$\sfrac{1}{\sqrt{6}} \left[(us + su) u - 2 uu s \right]$ 
\\
$\Sigma^0$ &
$\sfrac{1}{2}
\left[ (dsu+usd) -s (ud+du)
\right]$ 
& &
$\sfrac{1}{\sqrt{12}} \left[
s (du+ud) +(dsu+usd) -2(ud+du)s
\right]$ \\
$\Sigma^-$ &   $\sfrac{1}{\sqrt{2}} (ds -sd) d$  & &
$\sfrac{1}{\sqrt{6}}\left[ (sd + ds) d - 2 dd s \right]$  \\[.3cm]
$\Xi^0$ &     $\sfrac{1}{\sqrt{2}} (us -su) s$  & &
 $-\sfrac{1}{\sqrt{6}} \left[(ud + du) s - 2 ss u\right]$  \\
$\Xi^-$ &  
$\sfrac{1}{\sqrt{2}} (ds -sd) s$  & &
$-\sfrac{1}{\sqrt{6}} \left[(ds + sd) s - 2 ss d\right]$ \\
\hline
\hline
\end{tabular}
\end{center}
\caption{\footnotesize 
Mixed antisymmetric $\ket{M_A}$ 
and mixed symmetric $\ket{M_S}$ flavor states for the octet
baryons~\cite{Octet1,Octet2Decuplet}.}
\label{tabOctet}
\end{table*}

\subsection{Formalism}


In the covariant spectator quark model
the baryon wave functions $\Psi_B(P,k)$ 
depend on the baryon ($P$) and diquark ($k$) momenta,
as well as the flavor and spin projection indices.
Spin projection indices in the wave functions 
are suppressed for simplicity.

The wave functions of the octet baryon and 
the decuplet are constructed conveniently 
by the symmetrized states of the diquark (12),
and the off-mass-shell quark (3)~\cite{Nucleon,Nucleon2,Omega}.

The octet baryon $B$ wave functions can be expressed, 
in the $S$-wave approximation as~\cite{Octet2,Octet2Decuplet}
\ba
\Psi_B (P,k) = \frac{1}{\sqrt{2}}
\left[ \phi_S^0 \left| M_A \right>  + \phi_S^1 \left| M_S \right> 
 \right] \psi_B(P,k),
\ea
where $\phi_S^{0,1}$ are the spin-0 and spin-1 diquark 
components of the wave functions, 
$ \left| M_A \right>$ and $ \left| M_S \right>$ 
are the mixed antisymmetric and mixed symmetric flavor states,
and $\psi_B(P,k)$ is the radial wave function.
The explicit expressions for  $\phi_S^{0,1}$ are 
presented in Refs.~\cite{Octet2,Octet2Decuplet}.
The octet baryon flavor wave functions, are presented in Table~\ref{tabOctet}.

The decuplet baryon $B'$ wave functions,
in the $S$-wave approximation takes the form~\cite{Omega}
\ba
\Psi_{B'} (P,k) = - \psi_{B'} (P,k)  \left| B' \right>
\varepsilon_{P}^\alpha (\lambda) u_\alpha(P),
\ea
where $ u_\alpha(P)$ is the Rarita-Schwinger vector spin,
$\psi_{B'}(P,k)$ is the radial wave function, 
$\varepsilon_{P}^\alpha(\lambda)$ is the polarization state 
of the spin-1 diquark (polarization $\lambda)$, 
and $\left| B' \right>$ is the flavor wave function,
displayed in Table~\ref{tabDecuplet}.
For a more detailed description of the 
polarization states $\varepsilon_{P}^\alpha(\lambda)$
check Refs.~\cite{Nucleon,NDelta,NDeltaD}.

The radial wave functions $\psi_B(P,k)$ can be parametrized 
in terms of the variable
\ba
\chi_B = \frac{(M_B -m_B)^2 -(P-k)^2}{M_B m_D}.
\ea
The representation of $\psi_B(P,k)$ in 
terms of the single variable $\chi_B$ is possible because 
the baryon $B$ and the diquark are both on-mass-shell~\cite{Nucleon}.

The $\gamma^\ast B \to B^\prime$ transition current 
in relativistic impulse approximation 
takes the form~\cite{Nucleon,Nucleon2,Omega}
\ba
J^\mu = 3 \sum_{\Gamma} \int_k \overline \Psi_{B'} (P_+,k) j_q^\mu \Psi_B(P_-,k),
\ea
where $P_+$ ($P_-$) is the final (initial) baryon momentum, 
$k$ is the diquark momentum (on-mass-shell), 
and $j_q^\mu (q^2)$ is the quark current operator, 
depending on momentum transfer $q= P_+-P_-$~\cite{Nucleon,NDelta,NSTAR17}.
The integration symbol represents the covariant integration in $k$,
and the sum is over the diquark polarization states,
including the scalar and vector components.
The factor 3 takes into account the sum in the quarks 
based on the wave function symmetries.

The quark current $j_q^\mu$, where $q=u,d,s$,  
includes the electromagnetic structure 
of the constituent quark 
(gluon and quark-antiquark dressing effects)~\cite{Nucleon,Omega}.
The quark current operator is represented 
in the form~\cite{Omega} 
\ba
j_q^\mu (q) = j_1 \gamma^\mu + j_2 \frac{i \sigma^{\mu \nu} q_\nu}{2 M_N},
\ea
where $j_i$ ($i=1,2$) are the Dirac and Pauli flavor operators,
acting on the third quark component of the wave function, 
and $M_N$ is the nucleon mass, as before.


The components of the quark current $j_i$ ($i=1,2$) 
can be decomposed as the sum of operators 
\ba
j_i(Q^2)=
\sfrac{1}{6} f_{i+} (Q^2)\lambda_0
+  \sfrac{1}{2}f_{i-} (Q^2) \lambda_3
+ \sfrac{1}{6} f_{i0} (Q^2)\lambda_s,
\nonumber \\
\label{eqJq}
\ea
where
\ba
&\lambda_0=\left(\begin{array}{ccc} 1&0 &0\cr 0 & 1 & 0 \cr
0 & 0 & 0 \cr
\end{array}\right), \hspace{.3cm}
&\lambda_3=\left(\begin{array}{ccc} 1& 0 &0\cr 0 & -1 & 0 \cr
0 & 0 & 0 \cr
\end{array}\right),
\nonumber \\
&\lambda_s = \left(\begin{array}{ccc} 0&0 &0\cr 0 & 0 & 0 \cr
0 & 0 & -2 \cr
\end{array}
\right),
\label{eqL1L3}
\ea
are the flavor operators.
These operators act on the quark wave function in flavor space,
$q=  (\begin{array}{c c c} \! u \, d \, s \!\cr
\end{array} )^T$.

\begin{table}[t]
\begin{tabular}{l  c  c }
\hline
\hline
$B'$   & &  $\ket{B'}$      \\
\hline
\hline
$\Delta^+$  && $\sfrac{1}{\sqrt{3}}\left[uud + udu + duu  \right]$
\\
$\Delta^0$  && $\sfrac{1}{\sqrt{3}}\left[ddu + dud + udd  \right]$  \\[.3cm]
$\Sigma^{\ast +}$ &&  $\sfrac{1}{\sqrt{3}} \left[uus  + usu + suu  \right]$
\\
$\Sigma^{\ast 0}$ &&
$\sfrac{1}{\sqrt{6}} \left[uds + dus +usd +  sud + dsu + sdu  \right]$
\\
$\Sigma^{\ast -}$ &&
$\sfrac{1}{\sqrt{3}} \left[dds + dsd + sdd  \right]$ \\[.3cm]
$\Xi^{\ast 0}$ & $\;$& $\sfrac{1}{\sqrt{3}} \left[uss  + sus + ssu  \right]$
\\ 
$\Xi^{\ast -}$ && $\sfrac{1}{\sqrt{3}} \left[dss  + sds + ssd  \right]$
\\
\hline
\hline
\end{tabular}
\caption{\footnotesize
Quark flavor wave functions $\left|B'\right>$ for the decuplet
baryons~\cite{Omega}.
Not included here are the $\Delta^{++}$, $\Delta^-$ 
and $\Omega^-$ states.}
\label{tabDecuplet}
\end{table}

The functions  $f_{i+}$, $f_{i-}$ ($i=1,2$) 
represent the quark isoscalar and isovector 
form factors, respectively, based on 
the combinations of the quarks $u$ and $d$.
The functions $f_{i0}$ ($i=1,2$) represent 
the structure associated with the strange quark.

The explicit form for the quark form factors is 
included in Appendix~\ref{appQuarkFF}.
For the present discussion, the relevant part  
is that the quark form factors are represented 
in terms the vector meson mass poles 
associated with the mesons $\rho$,  $\omega$ 
and $\phi$ depending of the type ($l=\pm,0$).
The expressions of the quarks form factors 
are valid for the spacelike and timelike regions.
In the timelike region, however, the vector mass poles 
are corrected by finite decay widths.
The isovector transitions, like $\gamma^\ast N \to \Delta(1232)$
and $\gamma^\ast \Lambda(1116) \to \Sigma^0(1385)$, 
depend on the isovector form factors (meson $\rho$).
Other transitions depend on a combination 
of isovector, isoscalar and strange quark form factors.

Even though our quarks have structure, including processes 
which can be interpreted 
as meson cloud dressing of the quarks, 
there are processes involving
the meson cloud dressing that are not taken 
explicitly into account.
The processes in which there is a  meson exchange between the different
quarks cannot be represented by the quark dressing due to the meson cloud. 
Instead, the processes in which the meson is
exchanged between different quarks are regarded in our model,
as the meson is emitted and absorbed by baryon states, 
based on a baryon-meson molecular 
picture~\cite{Octet2Decuplet,DecupletDecays,Timelike2}.
Those effects are discussed in more detail 
in Sec.~\ref{secPionCloud}.

We consider here the covariant spectator 
quark model for the $\gamma^\ast B \to  B^\prime$ transition
from Refs.~\cite{Octet2Decuplet,DecupletDecays,Octet2}. 
As mentioned, we assume in first approximation that 
the octet baryon ($\Psi_B$) and the 
decuplet baryon  ($\Psi_{B'}$) wave functions 
are both described by the dominant $S$-wave 
quark-diquark configuration.
In the transition, only the symmetric flavor components 
of the octet baryon wave functions  ($\left| M_S \right>$)
contribute to the transition form factors,
because the decuplet baryon has no contributions from scalar diquarks.
The explicit expressions are presented in 
Refs.~\cite{Octet1,Omega,Octet2Decuplet,Octet2}.
In the $S$-wave approximation, the transition is dominated by 
the magnetic dipole form factor, $G_M$
as in the $\gamma^\ast N \to \Delta(1232)$ transition
($G_E= G_C \equiv 0$).
As a consequence, in Eq.~(\ref{eqGammaG}) 
we can replace $|G_T(q^2,W)|$ by $|G_M(q^2,W)|$.

When we take into account the pion cloud effects,
one can decompose $G_M$ into two components~\cite{NDelta,Lattice,Timelike2}
\ba
G_M(q^2; W) = 
G_M^{\rm B} (q^2,W) + G_M^{\pi}(q^2;W),
\label{eqGMtotal}
\ea
where $G_M^{\rm B}$ represent the contribution 
from the three-quark core (bare contribution) and $G_M^\pi$ represent 
the contribution from the pion cloud.
In the previous equation, we use $q^2= -Q^2$ 
to convert the spacelike relations for $G_M^{\rm B}$ and $G_M^\pi$ 
to the timelike region, and use $W$ to generalize the dependence 
of the form factor on the resonance mass 
($M_{B'}$ in the spacelike expressions).
We omit the indices $B$ and $B^\prime$ 
in the form factors for simplicity.
In some octet baryon to decuplet baryon electromagnetic transitions, 
the contributions of the kaon cloud may be also considered.
For a discussion of the magnitude of the kaon cloud 
contributions check Ref.~\cite{DecupletDecays}.

It is worth noticing that the dominance of 
the magnetic dipole form factor is an approximation,
and a consequence of the $S$-wave quark-diquark structure.
In the case of the $\gamma^\ast N \to \Delta(1232)$ transition
there is evidence that the quadrupole form factors 
$G_E$ and $G_C$ may have significant pion cloud 
contributions~\cite{GlobalFit,Siegert1}.
The contributions of those form factors to $|G_T( q^2,W)|$ 
from Eq.~(\ref{eqGT}) are, however, not significant,
since  $G_E$ is very small  
and $G_C$ is suppressed for small $q^2$.


The valence quark contribution $G_M^{\rm B}$
and the  pion cloud contribution $G_M^{\pi}$  are discussed 
in the two next subsections.
The numerical results for transition form factors are 
presented afterwards.
We anticipate here that as in the case of the 
$\gamma^\ast N \to \Delta(1232)$ transition,
the pion cloud/meson cloud contributions are relevant for the description 
of the  $\gamma^\ast B \to  B^\prime$ transitions.

\subsection{Valence quark contributions}
\label{secQuarks}

The contributions from the valence quarks to the 
octet baryon to decuplet baryon electromagnetic form factors 
($\gamma^\ast B \to B^\prime$) 
were calculated in previous works.
The expression for the magnetic form factor 
can be written as~\cite{Octet2Decuplet}
\ba
G_M^{\rm B} (q^2,W) = 
\frac{4}{3 \sqrt{3}} 
\, g_v \, {\cal I} (q^2,W),
\label{eqGMbare}
\ea 
where 
\ba
{\cal I} (q^2,W)= \int_k \psi_{B'} (P_+, k)  \psi_{B} (P_-, k),
\label{eqIntBBp}
\ea
is the overlap integral of the octet baryon and 
decuplet baryon radial wave functions, and 
\ba
g_v = \frac{1}{\sqrt{2}} 
\left[ \frac{2 M_B}{W  + M_B}
j_1^S (q^2) + \frac{M_B}{M_N} j_2^S (q^2) \right].
\ea 
The functions $j_i^S$ represent 
the projection of the flavor operators 
into the flavor components of the decuplet baryon 
and the mixed symmetric component 
of the octet baryon flavor state~\cite{Octet2Decuplet}.
The explicit expressions in terms of the quark 
form factors are presented in Table~\ref{tableJS}.

\begin{table}[t]
\begin{tabular}{l  c  c }
\hline
\hline
    & &  $j_i^S$      \\
\hline
\hline
$\gamma^* N \to \Delta$  && $\sqrt{2}f_{i-}$
\\[.5cm]    
$\gamma^* \Lambda \to \Sigma^{\ast 0}$ &&
$\sqrt{\frac{3}{2}} f_{i-}$
\\[.5cm] 
$\gamma^* \Sigma^+ \to \Sigma^{\ast +}$  &&
$\frac{\sqrt{2}}{6}(f_{i+} + 3 f_{i-} + 2 f_{i0})$
\\
$\gamma^* \Sigma^0 \to \Sigma^{\ast 0}$  &&
$ \frac{\sqrt{2}}{6}(f_{i+} + 2 f_{i0})$
\\
$\gamma^* \Sigma^- \to \Sigma^{\ast -}$ &&
$\frac{\sqrt{2}}{6}(f_{i+} - 3 f_{i-} + 2 f_{i0})$  \\[.5cm]    
$\gamma^* \Xi^0 \to \Xi^{\ast 0}$ &&
$\frac{\sqrt{2}}{6}(f_{i+} + 3 f_{i-} + 2 f_{i0})$
\\
$\gamma^* \Xi^- \to \Xi^{\ast -}$ &&
$\frac{\sqrt{2} }{6}(f_{i+} - 3 f_{i-} + 2 f_{i0})$
\\[0.02in]
\hline
\hline
\end{tabular}
\caption{\footnotesize
Coefficients $j_i^S$ ($i=1,2$) used to calculate the 
valence quark contributions for the transition form factors.
The label $\gamma^\ast N \to \Delta$ includes the $\gamma^\ast p \to \Delta^+$
and $\gamma^\ast n \to \Delta^0$ transitions ($n$ is the neutron).}
\label{tableJS}
\end{table}

In Table~\ref{tableJS} and along the draft, we use the asterisk ($^*$) 
to represent the excited states of $\Sigma$ and $\Xi$, 
members of the baryon decuplet.
The label $\gamma^\ast N \to \Delta$ includes the $\gamma^\ast p \to \Delta^+$
and $\gamma^\ast n \to \Delta^0$ transitions ($n$ is the neutron).

The overlap integral (\ref{eqIntBBp}) is invariant and 
can be evaluated in any frame.
For convenience we use the baryon $B'$ rest frame,
where $P_+= (W, {\bf 0})$,  $P_-= (E_B, - {\bf q })$,
with $E_B= \sqrt{M_B^2 + {\bf q}^2}$.
The momentum transfer takes the form 
\mbox{$q= (\omega, {\bf q })$}, where 
\ba
\omega = \frac{W^2 - M_B^2 + q^2}{2 W},
\hspace{1.cm}
 |{\bf q}| = \frac{\sqrt{y_+ y_-}}{2 W}.
\ea
The spacelike region, $q^2 \le 0$, is characterized by  
 \mbox{$|{\bf q}| \ge |{\bf q}|_0$} 
and the timelike region, $(W-M_B)^2 \ge q^2 > 0$, is  characterized 
by $0 \le  |{\bf q}| < |{\bf q}|_0$,
where  $|{\bf q}|_0 = \frac{W^2 - M_B^2}{2 W}$.

In the calculations, we use the experimental masses $M_N =0.939$ GeV,
$M_\Lambda =1.116$ GeV, $M_\Sigma = 1.192$ GeV and 
$M_\Xi = 1.318$ GeV, for the octet baryons.
As before, $W$ represents the decuplet baryon masses.
In the calculations associated with the physical decuplet baryons,
we use the physical masses: $M_\Delta = 1.232$ GeV, $M_{\Sigma^\ast} =1.385$ GeV, 
and $M_{\Xi^\ast} =1.533$ GeV.

The octet baryon radial wave functions 
take the form proposed on Refs.~\cite{Octet2,Octet1} 
for the study of the octet baryon electromagnetic form factors
\ba
\psi_N(P,k) =
\frac{N_N}{m_D (\beta_1 + \chi_N)(\beta_2 + \chi_N)}, \\
\psi_\Lambda(P,k) =
\frac{N_\Lambda}{m_D (\beta_1 + \chi_\Lambda)(\beta_3 + \chi_\Lambda)}, \\
\psi_\Sigma(P,k) =
\frac{N_\Sigma}{m_D (\beta_1 + \chi_\Sigma)(\beta_3 + \chi_\Sigma)}, \\
\psi_\Xi(P,k) =
\frac{N_\Xi}{m_D (\beta_1 + \chi_\Xi)(\beta_4 + \chi_\Xi)}, 
\ea 
where $N_B$ are normalization constants 
and $\beta_i$ ($i=1,2,3,4$) are 
square momentum-range parameters in units $M_B m_D$.
The parameters determined in Ref.~\cite{Octet1},
are $\beta_1= 0.0532$, $\beta_2= 0.809$,
 $\beta_2= 0.603$ and $\beta_2= 0.381$.
This parametrization reflects the natural 
order for the size of the baryon cores
$\beta_2 > \beta_3 > \beta_4$.

As for the decuplet baryon, we use 
the parametrization from Ref.~\cite{Omega}
\ba
& &
\psi_\Delta(P,k) = \frac{N_\Delta}{m_D(\alpha_1 + \chi_\Delta)^3}, \\
& &
\psi_{\Sigma^\ast}(P,k) = \frac{N_{\Sigma^\ast}}{
m_D(\alpha_1 + \chi_{\Sigma^\ast})^2 (\alpha_2 + \chi_{\Sigma^\ast})}, \\
& &
\psi_{\Xi^\ast}(P,k) = \frac{N_{\Xi^\ast}}{
m_D( \alpha_1 + \chi_{\Xi^\ast} ) ( \alpha_2 + \chi_{\Xi^\ast} )^2},
\ea
where $N_{B'}$ are normalization constants 
and  $\alpha_i$ ($i=1,2$) are square
momentum-range parameters in units $M_B m_D$.
In the present case the power associated with the factors 
in $\alpha_1$ and  $\alpha_2$ is related with the 
number of strange quarks (0, 1 or 2).
The radial wave function of the $\Omega^-$,
unnecessary for the present study, can be found in Ref.~\cite{Omega}.
In the calculations we use the values determined 
in the study of the decuplet baryon 
electromagnetic form factors 
$\alpha_1= 0.3366$ and $\alpha_2= 0.1630$~\cite{Omega}.

The normalization constants are determined by 
the conditions
\ba
\int_k \left[\psi_{B}(P,k) \right]^2 &= & 1, 
\nonumber \\
 \int_k \left[\psi_{B'}(P,k) \right]^2 &= &1.
\label{eqNormalizaN}
\ea
We consider positive values for all normalization constants.
The signs of the transition form factors are consequence 
of these conventions.

The octet baryon ($\psi_B$) and decuplet baryon ($\psi_{B'}$) 
radial wave functions, presented above, ensure that 
the valence quark contribution  
to $G_M$ defined by Eq.~(\ref{eqGMbare}) 
is proportional to $1/Q^4$ for very large $Q^2$~\cite{NDelta},
consistent with estimates from perturbative QCD (pQCD)~\cite{Carlson}.

The parametrizations of the octet baryon and decuplet baryon 
radial wave functions were obtained from fits to the 
lattice QCD simulations of the  
electromagnetic form factors 
for pion masses larger than 350 MeV 
(small meson cloud contributions)~\cite{Octet2,Omega,Lin09,Boinepalli09} .
The estimates of the valence quark contributions to the 
octet baryon and decuplet baryon elastic form factors 
are extrapolated to the physical regime 
using our extension of the model from the lattice to the physical case.
Details of the procedure can be found 
in Refs.~\cite{Lattice,LatticeD,Octet1,Octet2,Omega}.

Our estimates for the $\gamma^\ast N \to \Delta(1232)$ 
transition form factors compare very well 
with the lattice QCD simulations with the 
corresponding pion masses~\cite{LatticeD}.
Our results are also consistent with 
the bare core estimates from the 
EBAC model~\cite{LatticeD,Timelike2}.
The EBAC model is a meson-baryon coupled-channel 
dynamical model where the meson-baryon couplings are calibrated by 
the pion electro-production data and photo-production data~\cite{Burkert04,JDiaz07}.
The contributions of the bare core are obtained 
when we set the meson-baryon coupling to zero~\cite{JDiaz07}.

Based on the results for the $\gamma^\ast N \to \Delta(1232)$ 
for the lattice QCD regime,  
where meson cloud effects are negligible,
and on the comparison with the EBAC results at the physical point, 
one can conclude that the calibration 
of the valence quark degrees 
of freedom is under control~\cite{DecupletDecays}.
Our parametrizations of the pion cloud contributions,
discussed below, are inferred from the comparison between 
the extrapolation to the physical limit
and the physical data~\cite{NDeltaD,Timelike2,NSTAR17}.

A final note about the global normalization 
of the wave functions is in order.
The wave functions associated to the baryon decuplet
are normalized properly because the decuplet baryons 
are described by a model where we neglect the pion cloud contributions.
As for the baryon octet, the normalization
of the valence quark component is modified due 
to the inclusion of the pion cloud component.
We note, however, that this correction 
only affects $G_M^{\rm B}$ and that, 
due to the magnitude of the normalization constant 
and the relative contribution from the valence quark contributions, 
the normalization effects can be estimated as 3\% at most.
One concludes, then, that in a first approximation,
we can ignore the normalization correction due to the pion cloud dressing.



\subsection{Pion cloud contributions}
\label{secPionCloud}


The pion cloud contribution to the 
$\gamma^\ast B^\prime \to B$ transition 
are estimated by the $SU(3)$ extension of 
our pion cloud model for the 
$\gamma^\ast N \to \Delta(1232)$ transition~\cite{NDelta,NDeltaD,LatticeD}.

\begin{figure}[t]
\vspace{.4cm}
\includegraphics[width=2.7in]{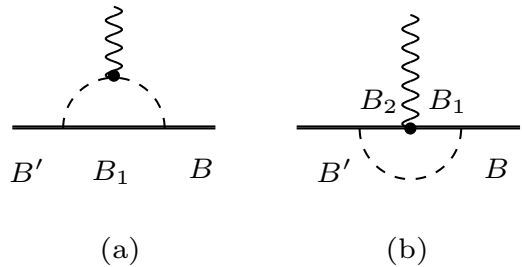}
\caption{\footnotesize
Meson cloud contributions
for the electromagnetic transition form factors.
Between the initial octet ($B$) and
final decuplet ($B'$) baryon states,
there are several possible intermediate baryon states:
$B_1$ in diagram (a); $B_1$ and $B_2$ in diagram (b).
}
\label{figMesonCloud}
\end{figure}

We use, in particular the results of Ref.~\cite{DecupletDecays},
where the meson cloud contributions of the diagrams of Fig.~\ref{figMesonCloud} 
are determined explicitly in the limit $q^2=0$.
The calculations of the meson cloud loops 
are based on the cloudy bag model~\cite{Thomas84,Theberge83,Yamaguchi89}.
The explicit calculations use the meson-baryon couplings  
for the possible octet baryon and decuplet baryon 
intermediate states from Fig.~\ref{figMesonCloud}.
The connection with the quark microscopic properties  
between the covariant spectator quark model and the cloudy bag model 
is performed matching the Dirac and Pauli couplings.
In Ref.~\cite{DecupletDecays},
in addition to the pion, we considered also 
the contributions of the kaon and the eta~\cite{DecupletDecays}.
The eta contributions prove to be very small.
More details about the meson and baryon 
contributions to the processes from Fig.~\ref{figMesonCloud}
are included in Appendix~\ref{appMC}.

In the present work, we consider the simplest 
approximation, taking into account only the pion cloud contributions,
and drop the kaon cloud contributions,
since the extrapolation of the pion cloud contributions to 
finite $q^2$, based on the results of the 
$\gamma^\ast N \to \Delta(1232)$ transition is straightforward.

The generalization of the pion cloud contributions 
to the timelike region follows the lines 
of our work for the  $\gamma^\ast N \to \Delta(1232)$~\cite{Timelike2}.
We represent then 
\ba
G_M^\pi (q^2)& =& 
G_M^{\pi {\rm a}} (0) \, F_\pi (q^2) 
\left(  \frac{\Lambda_\pi^2}{\Lambda_\pi^2 - q^2} \right)^2 
\nonumber \\
& & + \, G_M^{\pi {\rm b}} (0)  \, \tilde G_D^2 (q^2), 
\label{eqGMpi}
\ea
where $G_M^{\pi {\rm a}} (0)$ 
and $G_M^{\pi {\rm b}} (0)$ are the pion contributions
for the diagrams (a) and (b)
in the limit $q^2=0$, respectively, 
$F_\pi$  is the pion electromagnetic form factor, $\Lambda_\pi^2 = 1.53$ GeV$^2$,
and $\tilde G_D$ is a generalization 
of the traditional dipole form factor.
The coefficients  $G_M^{\pi {\rm a}} (0)$ 
and $G_M^{\pi {\rm b}} (0)$ are presented in Table~\ref{tablePionCloud}.
In Eq.~(\ref{eqGMpi}) we omit the dependence on $W$, 
since the coefficients  $G_M^{\pi {\rm a}} (0)$ 
and $G_M^{\pi {\rm b}} (0)$ are determined in the physical limit ($W=M_{B'}$).

We use the parametrization~\cite{Timelike2}
\ba
F_\pi (q^2) = \frac{\alpha}{\alpha  - q^2 -
\frac{1}{\pi}\beta q^2 \log \frac{q^2}{m_\pi^2} + i \beta q^2},
\label{eqFpi}
\ea 
where 
$\alpha = 0.696$ GeV$^2$, $\beta = 0.178$
and $m_\pi$ is the mass of the pion.
In the spacelike region $F_\pi$ takes the form
(analytic continuation)
\ba
F_\pi (q^2) = \frac{\alpha}{\alpha  - q^2 -
\frac{1}{\pi} \beta q^2 \log \frac{(-q^2)}{m_\pi^2} }.
\ea

Equation (\ref{eqFpi}) is derived from
an analytic expression which include the structure 
of the two-pion threshold~\cite{Iachello04,Timelike,Timelike2}
for $q^2 \gg 4 m_\pi^2$, 
in order to obtain a simpler parametrization of the $F_\pi$ data.  
Although the two-pion structure is not included explicitly,
the error in the approximation is small, since 
the imaginary component has a small magnitude 
in the region $0 \le q^2 \le 4 m_\pi^2$.
One derives, then a smoother approximation 
to the imaginary part of $F_\pi$ without significant 
loss of accuracy.
Higher precision parametrizations based on more 
complex analytic structures and a larger 
number of parameters can be found in Refs.~\cite{Donges95,Hanhart12,Herrmann93}.

Following Ref.~\cite{Timelike2}, 
the function $\tilde G_D$ is defined as
\ba
\tilde G_D (q^2) =
\frac{\Lambda_D^4}{(\Lambda_D^2 - q^2)^2 + \Lambda_D ^2 \Gamma_D^2}, 
\label{eqGD}
\ea
where $\Lambda_D^2 = 0.9$ GeV$^2$ and 
$\Gamma_D (q^2)$ is an effective width.  
The explicit expression for $\Gamma_D (q^2)$ 
is presented in Appendix~\ref{appRegularization}.

The parametrization of (\ref{eqGMpi}) is motivated by 
the fast suppression of the pion cloud contributions
in the spacelike region.
This effect is simulated by simple multipole functions,
and with the direct photon coupling with the pion
in the diagram \ref{figMesonCloud}(a).
The second term simulates the contributions 
from the diagram \ref{figMesonCloud}(b) and therefore includes 
the contributions from several
intermediate electromagnetic transitions between octet 
and/or decuplet baryon states (check Appendix~\ref{appMC}).
The multipole powers are chosen using the 
expected falloff for large $Q^2$, 
estimated by pQCD~\cite{Carlson}.
Analysis based on pQCD suggests that the 
valence quark contributions dominate $G_M$ 
and that $G_M \propto 1/Q^4$.
Extending the analysis for the meson cloud effects,
interpreted as the contributions of meson-baryon systems, 
one concludes that those contributions\footnote{Using pQCD 
one can show that the leading order form factor  
with $n$ active constituents behaves for large $Q^2$
like $1/Q^{2(n-2)}$. 
For a system of three quark, one obtains the 
a falloff with $1/Q^4$. 
For a system of three quarks and a quark-antiquark pair
(5 constituents), 
resembling a baryon-meson system, 
one expect then a falloff with $1/Q^8$.
Meson cloud contributions are then characterized by an 
extra suppression of $1/Q^4$ for large $Q^2$.}
are ruled at very large $Q^2$ by $G_M \propto 1/Q^8$.
The second term of  (\ref{eqGMpi}) falls off with $ 1/Q^8$.
The first term of  (\ref{eqGMpi}) 
falls of with $1/(Q^6 \log Q^2)$, still close to the expected rule.

\begin{table}[t]
\begin{tabular}{l  r r r c}
\hline
\hline
       &  $G_M^{\pi {\rm a}} (0)$  & $G_M^{\pi {\rm b}} (0)$ &  $G_M^{\pi} (0)$ &
$G_M^{\rm B} (0,M_{B'})$\\
\hline
\hline
$\gamma^\ast N \to \Delta$ & 0.713 & 0.610 &  1.323 & 1.633\\ [.2cm]
$\gamma^\ast \Lambda \to \Sigma^{* 0}$  & 0.669 & 0.358 &  1.027 & 1.683\\ [.2cm]
$\gamma^\ast \Sigma^+ \to \Sigma^{* +}$ & 0.149 & 0.513 &  0.663 & 2.094\\
$\gamma^\ast \Sigma^0 \to \Sigma^{* 0}$ & 0.000 & 0.270 &  0.270 & 0.969\\
$\gamma^\ast \Sigma^- \to \Sigma^{* -}$ & $-$0.149 & 0.026 &  $-$0.124 & $-$0.156 \; \\[.2cm]
$\gamma^\ast \Xi^0 \to \Xi^{\ast 0}$  & 0.222 &  0.086 &  0.308 & 2.191\\
$\gamma^\ast \Xi^- \to \Xi^{\ast -}$  & $-$0.222 & 0.084  & $-$0.138 & $-$0.168 \; \\
\hline
\hline
\end{tabular}
\caption{\footnotesize
Coefficients of the pion cloud contributions.
In the last column, we include the bare contribution at $q^2=0$.
}
\label{tablePionCloud}
\end{table}

The extension of the model with the inclusion of the kaon cloud 
will require the generalization of the two terms from 
Eq.~(\ref{eqGMpi}) to the case of the kaon.
This non trivial generalization is planed for a future work.


In the last column of Table~\ref{tablePionCloud}, 
we include for convenience the bare contribution $G_M^{\rm B}(0, M_{B'})$
to the magnetic form factor at $q^2=0$.
The relative magnitude of the pion cloud contribution at $q^2=0$ can 
then be estimated by $G_M^{\pi}(0)/(G_M^{\rm B}(0, M_{B'}) + G_M^{\pi}(0))$.

\begin{figure*}[t]
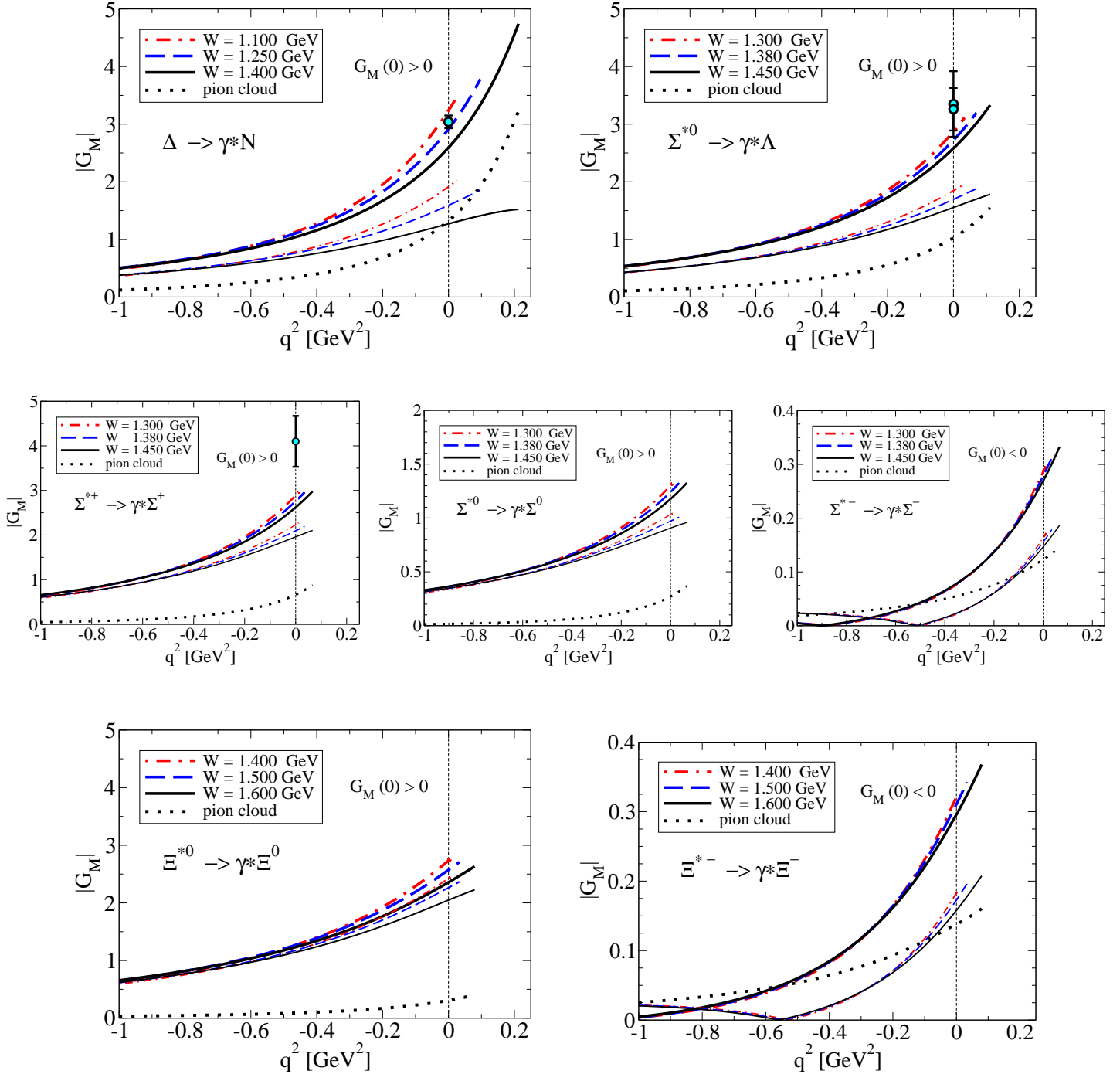

\centerline{\vspace{0.5cm}  }
\centerline{
\mbox{
\includegraphics[width=3.1in]{GM-Nucleon-v3b} \hspace{.6cm}
\includegraphics[width=3.1in]{GM-Lambda-v3b} }}
\vspace{.7cm}
\centerline{
\mbox{
\includegraphics[width=2.4in]{GM-SigmaP-v3b} \hspace{.1cm}
\includegraphics[width=2.4in]{GM-Sigma0-v3} \hspace{.1cm}
\includegraphics[width=2.4in]{GM-SigmaM-v3} 
}}
\vspace{.9cm}
\centerline{
\mbox{
\includegraphics[width=3.1in]{GM-Xi0-v3} \hspace{.6cm}
\includegraphics[width=3.1in]{GM-XiM-v3} }}
\caption{\footnotesize{
Magnitude of transition form factor $G_M$.
The thick lines represent the total
(valence plus pion cloud) and 
the thin lines represent the valence quark contribution.
The dotted line represent the pion cloud contributions
(follow the discussion in the main text).
The Data are from Table~\ref{tableGM0}.
}}
\label{figGM}
\end{figure*}

\subsection{Transition form factors}
\label{secFormFactors}

We now discuss the results 
for the transition form factor associated 
to Eqs.~(\ref{eqGMtotal}), (\ref{eqGMbare}) and (\ref{eqGMpi}).
Our transition form factors are real functions 
(by construction) in the spacelike region,
and became complex only in the timelike region ($q^2 > 0$).
We present the results for $|G_M|$, because 
only the magnitude of $G_M$ is relevant for the radiative and Dalitz decays.
The sign of $G_M$ in the spacelike region 
is the consequence of the our convention to the flavor states 
presented in Tables~\ref{tabOctet} and \ref{tabDecuplet}.

The numerical results for $|G_M|$, 
for several values of $W$ near the physical mass $M_{B'}$ 
are presented in Fig.~\ref{figGM} by the thick lines.
For the $\Sigma^\ast$ decays the 
we choose a range of variation based on the 
$\Sigma^\ast$ total decay width, 
and on the range of the HADES simulations~\cite{Lalik19}.
For the $\Xi^\ast$ decays, since the decay width 
is very small  (about 10 MeV), we consider a wider range 
for a better visualization of the dependence on $q^2$.

In addition to $|G_M|$, we include also 
the result of the valence quark contribution  $|G_M^{\rm B}|$
(thin lines) and the absolute values of the pion cloud 
contribution (dotted line), according to Eq.~(\ref{eqGMpi}).
The line associated to the pion cloud corresponds, 
in fact, to the estimate associated with the largest value of $W$.
The remaining cases have the same shape, except that
the estimates are limited to $q^2 \le (W - M_B)^2$.

The data included in the graph represent 
the magnitude of the experimental magnetic form factors for $q^2=0$,
estimated from the radiative decay width data.
The $G_M(0)$ data is discussed in more detail in the next section.
The experimental values for  $|G_M(0)|$ are important to infer the 
accuracy of the constant form factor model.
The model associated with the constant form factor 
corresponds to a horizontal line 
with the magnitude of the experimental value for $|G_M(0)|$.

One can notice that the model estimates 
for the $\Delta \to \gamma^\ast N$ and 
$\Sigma^{\ast 0} \to \gamma^\ast \Lambda$ decays
have a magnitude comparable with the data.
In the case of the  $\Sigma^{\ast +} \to \gamma^\ast \Sigma^+$ 
decay the model underestimate clearly the data.
This underestimation is in part the consequence of neglecting 
the kaon cloud contributions.
When those effects are taken into account one obtain 
$G_M(0) = 3.22$, only 1.5 standard deviations 
from below the data~\cite{DecupletDecays}.

In Fig~\ref{figGM}, one can observe the dependence 
of the transition form factors on the variable $W$.
In general, for a fixed value of $q^2$ the magnitude 
of $G_M$ decreases with $W$, 
as a consequence of our analytic expressions for $G_M^{\rm B}$.
This $W$-dependence was tested in our calculations 
in the lattice QCD regime, where the masses 
of the baryons and mesons are larger that the physical ones~\cite{NSTAR17,Lattice,LatticeD}.
The $W$-dependence of our results is an important 
characteristic of our formalism, which has an impact 
on the calculation of the radiative and the Dalitz decay 
widths in terms of $W$, presented in the Sec.~\ref{secGammaW}.

In this aspect the present model is distinct of 
other models, like the constant form factor model 
and some VMD models~\cite{Krivoruchenko02}.
The Iachello-Wan model~\cite{Iachello04,Dohrmann10} 
includes only a weak $W$-dependence on the transition form factors.


In the graphs, the spacelike results for $G_M$ are equivalent 
to the results presented in the graph for $|G_M|$,
in most cases, since $G_M(0) > 0$.
The exceptions are the $\Sigma^{* -} \to \gamma^\ast \Sigma^-$
and $\Xi^{* -} \to \gamma^\ast \Xi^-$ decays, where $G_M(0) < 0$, 
according to the estimates from Ref.~\cite{DecupletDecays}.
Our numerical values for $G_M(0)$ are presented 
in the next section (see Table~\ref{tableGM0}).

The results for the $\Delta \to \gamma^\ast N$ form factors 
are almost identical to the results from Ref.~\cite{Timelike2},
except that in the previous work we use the 
approximation $G_M^{\pi {\rm a}}(0) = G_M^{\pi {\rm b}}(0) =\frac{1}{2} G_M^{\pi}(0)$
(pion cloud contributions equally divided 
between the two pion cloud processes 
from Fig.~\ref{figMesonCloud}).
The results for the $\Delta \to \gamma^\ast N$ form factors are interesting 
because there is a deeper penetration 
in the timelike region due the large 
values of the upper limit $(W-M_B)^2$,
where $B$ is the nucleon.
For larger values of $W$ one can notice that the valence 
quark contribution line became more flat.
In that region, one can also observe   
the enhancement of $|G_M|$ for large $q^2$, 
a direct consequence of the pion cloud contribution 
regulated by Eqs.~(\ref{eqGMpi}) and (\ref{eqFpi}),
characterized by the peak of $F_\pi (q^2)$ near 
$q^2 \approx m_\rho^2 \simeq 0.6$ GeV$^2$.

\begin{table*}[t]
\begin{center}
\begin{tabular}{l  cc  cc   cc  cc cc cc}
\hline
\hline
     &&  \sp\sp$G_M(0)$   &&  
\sp\sp$\left. G_M(0)\right|_\pi $  && $|G_M(0)|_{\rm exp}$  &&
$\Gamma ({\rm keV}) $ && $\Gamma_{\rm exp} ({\rm keV})$
\\ 
\hline
$\Delta \to \gamma N$ &&  3.02   && 2.96 &&
\sp\sp$3.04\pm0.11$ \cite{PDG10}
&& 648 && $660\pm47$ \cite{PDG10}  \\[.15cm]
$\Sigma^{\ast 0} \to \gamma \Lambda$ && 3.08   && 2.71 &&
\sp\sp$3.35\pm0.57$ \cite{PDG10} && 399 && $470\pm 160$ \cite{PDG10} \\
  &&    &&  &&
\sp\sp$3.26\pm0.37$ \cite{Keller12} &&  && $445\pm 102$
\cite{Keller12} \\[.15cm]
$\Sigma^{\ast +} \to \gamma \Sigma^+ $ && 3.22 && 2.76&&
\sp\sp$4.10\pm0.57$
\cite{Keller11a}  && 154 && $250\pm70$ \cite{Keller11a} \\
$\Sigma^{\ast 0} \to \gamma \Sigma^0 $ && 1.46 && 1.24 && $< 11$~\cite{Colas75} 
&& 32 && $< 1750$~\cite{Colas75}\\
$\Sigma^{\ast -} \to \gamma \Sigma^-$ && $-0.31$ \sp &&   $-0.28$\sp\sp
&&   $< 0.8$ \cite{Molchanov04} && 1.4 && $< 9.5$  \cite{Molchanov04} 
\\ [.15cm]
$\Xi^{\ast 0} \to  \gamma \Xi^0 $ && 3.29  && $2.50$ &&     && 182 &&      \\   
$\Xi^{\ast -} \to  \gamma \Xi^- $ && $-0.38$ \sp && $-0.31$\sp\sp &&  $< 4.2$ \cite{Ablikim19} && 2.4 &&   $< 366$ \cite{Ablikim19}\\
\hline
\hline
\end{tabular}
\end{center}
\caption{\footnotesize
Results for $G_M(0)$ corresponding to the $B' \to \gamma \, B$ decays. 
The values for $|G_M(0)|_{\rm exp}$ are estimated 
using the experimental values of $\Gamma_{B' \to \gamma B}$.
$\left. G_M(0)\right|_\pi $ is the estimate when 
we omit the kaon cloud contributions (only pion cloud).}
\label{tableGM0}
\end{table*}

The first detailed study of the $\Delta(1232)$ Dalitz decay   
at HADES suggests that the constant form factor model
is insufficient to describe the data and 
that the signature of the form factor dependence on $q^2$
is present in the data~\cite{HADES17}.

The present calculations also suggest that the constant form factor model 
is not a good approximation for the 
$\Sigma^{* 0} \to \gamma^\ast \Lambda$, $\Sigma^{* -} \to \gamma^\ast \Sigma^-$ and 
$\Xi^{* -} \to \gamma^\ast \Xi^-$ decays, 
since in those cases $|G_M|$ is significantly enhanced 
near the pseudothreshold. 
Those enhancements can be the consequence of 
the bare contribution ($\Sigma^{* -} \to \gamma^\ast \Sigma^-$ and 
$\Xi^{* -} \to \gamma^\ast \Xi^-$) 
or the pion cloud contribution \mbox{($\Sigma^{* 0} \to \gamma^\ast \Lambda$).}

The results for the $\Sigma^{* +} \to \gamma^\ast \Sigma^+$,
$\Sigma^{* 0} \to \gamma^\ast \Sigma^0$
and $\Xi^{* 0} \to \gamma^\ast \Xi^0$ transitions indicate 
that the relative pion cloud contributions 
are smaller than in the other transitions.

From the graphs for  $|G_M|$, we also conclude that there are different 
classes of magnitudes:
$\Delta \to \gamma^\ast N$ and $\Sigma^{* 0} \to \gamma^\ast \Lambda$;
 $\Sigma^{* +} \to \gamma^\ast \Sigma^+$ and $\Xi^{* 0} \to \gamma^\ast \Xi^0$ 
(large magnitude);
  $\Sigma^{* 0} \to \gamma^\ast \Sigma^0$ (moderate magnitude);
 $\Sigma^{* -} \to \gamma^\ast \Sigma^-$ and $\Xi^{* -} \to \gamma^\ast  \Xi^-$ 
(small magnitude)~\cite{Octet2Decuplet}.
The impact of these magnitudes on the Dalitz
decay widths is discussed in Sec.~\ref{secDalitzRates}.

In the graphs for the $\Sigma^{* -} \to \gamma^\ast \Sigma^-$
and $\Xi^{* -} \to \gamma^\ast \Xi^-$ transitions, 
one can notice nodes in both $|G_M|$ and $|G_M^{\rm B}|$
in the spacelike region.
Those nodes are a consequence of zeros of $G_M$ due to a sign change.
Since both form factors are negative near $q^2=0$,
the nodes indicate the point where the functions became negative.  
The zero crossings are the consequence 
of the sign change of the valence quark contributions,
according to our $SU(3)$ parametrization 
of the quark form factors (see Table~\ref{tableJS}).
Similar results were also obtained in a previous 
study based on the covariant spectator quark model~\cite{Octet2Decuplet},
with a not so general description of the pion cloud contributions.

The $\Sigma^{* -} \to \gamma^\ast \Sigma^-$
and $\Xi^{* -} \to \gamma^\ast \Xi^-$ transitions are 
the transitions with smaller valence quark contributions.
This result is also a consequence of our approximated $SU(3)$ symmetry.
In the exact $SU(3)$ limit the form factors 
$f_{i+}$, $f_{i-}$ and $f_{i s}$ are undistinguished 
and the valence quark contribution vanishes 
because $j_i^S \equiv 0$, according to Table~\ref{tableJS}. 
The small but non-zero contributions to $G_M^{\rm B}$ 
are then the consequences of a small $SU(3)$ symmetry breaking.

\subsection{Comparison with the literature}

Our estimates can be compared directly with other 
estimates based on valence quark degrees of freedom.

Calculations based on non relativistic and relativistic 
quark models~\cite{Kaxiras85,Darewych83,Sahoo95,Wagner98,Bijker00}
underestimate in general $G_M$ near $Q^2=0$,
which may be interpreted as a consequence 
of the absence of meson cloud effects.
Also lattice QCD simulations underestimate $G_M(0)$~\cite{Leinweber93}.
In Ref.~\cite{Octet2Decuplet}, we compare explicitly 
our upper limit for the valence quark contribution for $G_M(0)$, 
defined by Eq.~(\ref{eqGMbare}) with ${\cal I}(0,M_{B'})=1$, 
with the lattice results from Ref.~\cite{Leinweber93}.
We conclude that the two estimates are very close, 
within the lattice QCD uncertainties.

We now compare our estimates of the valence quark contributions 
with estimates based on the Dyson-Schwinger framework from Ref.~\cite{Alepuz18}, 
also based on the valence quark degrees of freedom.
Our results for $G_M^B$ compare well with the estimates from Ref.~\cite{Alepuz18} 
above \mbox{1 GeV$^2$,}
for transitions with larger magnitude for $|G_M|$,
suggesting that the two methods have similar 
predictions for the large-$Q^2$ region.
For the $\Sigma^{- \ast}$ and $\Xi^{- \ast}$ decays,
both formalisms predict small but different magnitudes.
Recall that those transitions are more sensitive 
to the mechanisms of $SU(3)$ symmetry breaking. 
Both formulations predict that the quark core 
contributions vanish in some point between 0 and 1 GeV$^2$.
Below \mbox{$Q^2=1$ GeV$^2$,} the comparison is more delicate,
because the Dyson-Schwinger estimates are presented as an interval of variation.
From the results for $Q^2=0.1$ and 0.2 GeV$^2$, 
one can conclude that we overestimate
the results from Ref.~\cite{Alepuz18}  in about 30\%--50\%.

The transition form factors have also been 
calculated with a $SU(3)$ chiral quark-soliton model~\cite{Kim20a},
taking into account some pion production from the nucleon.
The model explains well 
the $\gamma^\ast N \to \Delta(1232)$ lattice QCD data 
for $G_M$ for the corresponding pion mass.
The model calibrated by $Q^2 \simeq 0$ data describe 
well the low-$Q^2$ data but falls off slower that the 
experimental data.
The estimates of the reaming transition form factors 
compare well with our estimates of the bare contribution 
to $G_M(0)$ (see Table~\ref{tablePionCloud}),
but differ in sign.
The unnormalized estimate of  $G_M$  for 
the $\gamma^\ast N \to \Delta(1232)$ transition
is also similar to our estimate for $G_M^{\rm B}(0)$.
Their form factors have a slower falloff with $Q^2$
when compared with our estimates.

When we restrict the analysis to $Q^2=0$ there 
are several frameworks which provide estimates for $|G_M(0)|$
closer to the available data.
There are calculations based on chiral 
perturbation theory~\cite{Butler93a,Arndt04,Holmberg18} 
and the large $N_c$ limit~\cite{Lebed11}.
Those estimates are restricted in the range of $Q^2$,
and rely on the determination of low-energy constants.
Also calculations based on QCD sum rules 
predict large contributions to $|G_M(0)|$
in comparison with our estimates~\cite{Aliev06,Wang09}.
One notices, however, that those the comparison 
between quark models and QCD sum rules have to be 
performed with care, since the normalization 
in QCD sum rules is based on distribution amplitudes 
defined for large $Q^2$, in contrast with 
quark models, where the normalization is defined at $Q^2=0$.

In the next section, we study the impact 
of our model for the transition form factors on 
the radiative and Dalitz decay widths.



\section{Radiative and Dalitz decay widths}
\label{secDalitz}

We present here our estimates for the $B^\prime$
radiative and Dalitz decay widths.
We start with the radiative decays 
at the pole: $\Gamma_{\gamma B} (M_{B'})$.
Later on, we discuss the functions 
$\frac{d \;}{d q}  \Gamma_{e^+e^-B} (q,W)$,
$\Gamma_{e^+ e^- B} (W)$ and  $\Gamma_{\gamma B} (W)$.

\subsection{Electromagnetic decay widths}
\label{secRadiative}

Using the dominance of the magnetic dipole form factor, 
we can write~\cite{Krivoruchenko02,Octet1,Octet2}
\ba
\Gamma_{\gamma B} = \frac{\alpha}{16}
\frac{(M_{B'}^2 -M_B^2)^3}{M_{B'}^3 M_B^2} |G_M(0)|^2.
\label{eqGammaB2}
\ea
In Table~\ref{tableGM0}, we present the model estimates for $G_M(0)$ and $\Gamma \equiv \Gamma_{\gamma B}$
in the second and fifth columns,
and compare those estimates with 
the experimental data~\cite{PDG10,Keller11a,Colas75,Molchanov04,Ablikim19,PDG18}, 
in the fourth and sixth columns.
$|G_M(0)|_{\rm exp}$ is determined from  $\Gamma_{\rm exp}$ using  Eq.~(\ref{eqGammaB2}).
The numerical results were calculated in Ref.~\cite{DecupletDecays}.
For the decays for which there are no data,
we include the experimental estimate of the upper limit
when available.

The estimate of the third column, $\left. G_M(0) \right|_\pi$,
correspond to the calculation which exclude the 
kaon cloud contribution (only pion cloud), as in Sec.~\ref{secFormFactors}.  

As discussed in the previous section,
our estimate of $G_M(0)$, given by $\left. G_M(0) \right|_\pi$,
is consistent with the data for the 
$\Delta \to \gamma N$ and $\Sigma^{* +} \to \gamma \Lambda$ decays,
and underestimates the result for the $\Sigma^{* +} \to \gamma \Sigma^+$ decay. 
Calculations based on chiral perturbation theory~\cite{Butler93a,Holmberg18},
large $N_c$ limit~\cite{Lebed11} and QCD sum rules~\cite{Wang09}, 
compare well with the available data.
A detailed comparison between model estimates and 
experimental data can be found in Refs.~\cite{DecupletDecays,Octet2Decuplet}.

On Table~\ref{tableGM0}, one can notice that the experimental limit 
for the $\Sigma^{\ast -}$ decay is close to our model estimate.
One can conclude then that there is some hope that this 
decay width can be measured in a near future.

\begin{figure*}[t]
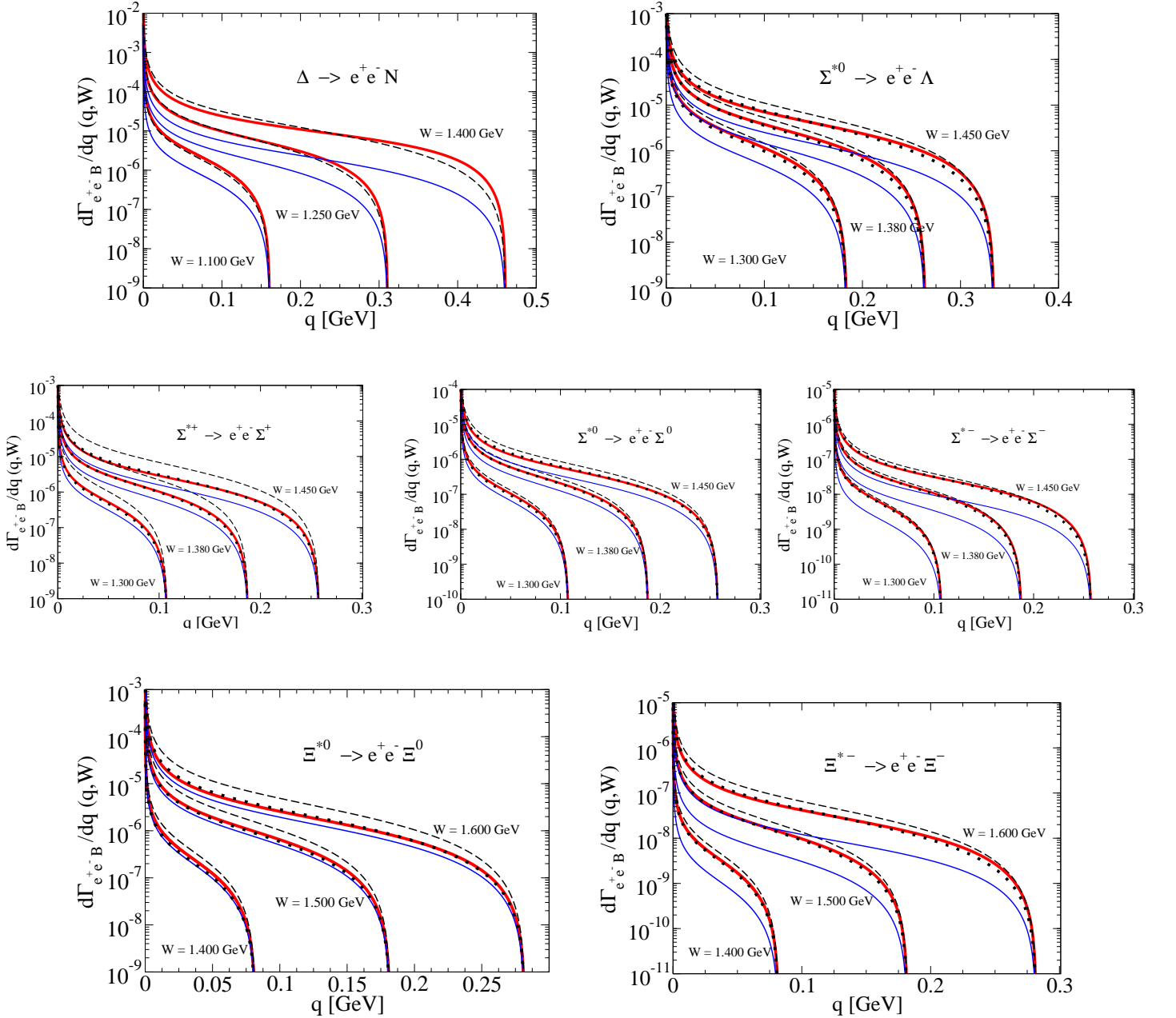

\centerline{\vspace{0.3cm}  }
\centerline{
\mbox{
\includegraphics[width=3.1in]{DGamma-Nucleon-v2} \hspace{.6cm}
\includegraphics[width=3.1in]{DGamma-Lambda-v3} }}
\vspace{.7cm}
\centerline{
\mbox{
\includegraphics[width=2.4in]{DGamma-SigmaP-v3} \hspace{.3cm}
\includegraphics[width=2.4in]{DGamma-Sigma0-v3} 
\includegraphics[width=2.4in]{DGamma-SigmaM-v3} 
}}
\vspace{.7cm}
\centerline{
\mbox{
\includegraphics[width=3.1in]{DGamma-Xi0-v3} \hspace{.6cm}
\includegraphics[width=3.1in]{DGamma-XiM-v3} }}
\caption{\footnotesize{Dalitz decay rates 
$\frac{d \;}{d q}  \Gamma_{e^+e^-B}$
for different values of $W$.
Note a difference of scales.
The thick solid lines represent our final estimation (bare plus pion cloud).
The thin solid lines represent the bare quark approximations.
The results of the constant form factor model 
($G_M(q^2) \to G_M(0)$) are indicated by the dashed lines.
The dotted lines represent the estimate of the 
constant form factor model when we exclude the kaon cloud 
($G_M(q^2) \to \left. G_M(0) \right|_\pi$).
}}
\label{figDGamma}
\end{figure*}


\subsection{Dalitz decay rates}
\label{secDalitzRates}

The results for the Dalitz decay rates are presented in Fig.~\ref{figDGamma},
for all the decuplet baryon decays, for several values of $W$.
We include the labels $B^\prime \to e^+ e^- B$ 
in order to identify the decaying decuplet baryon.
Recall that $\Sigma^0(1385)$ decay on  $\Lambda(1116)$
and on $\Sigma^0(1193)$.

The thick solid lines indicate the final result:
the combination of valence quark and pion cloud contributions.
The thin lines indicate the valence quark contributions
(when we drop the pion cloud contributions).

The dashed lines indicate the result 
of the constant form factor model, 
obtained when we consider:  $G_M \equiv G_M(0)$,
also known as QED estimate.
To represent the QED estimate, 
we consider the following convention:
\begin{itemize}
\item
In cases where experimental data exist  
($\Delta \to \gamma \, N$, $\Sigma^{\ast 0} \to \gamma \, \Lambda$ and
$\Sigma^{\ast +} \to \gamma \, \Sigma^+$),
we use the magnetic form factor determined by the electromagnetic
decay width (see Table~\ref{tableGM0}). 
For the $\Sigma^{\ast 0} \to \gamma \, \Lambda$ transition 
we approximate the result
by the central value ($|G_M(0)|_{\rm exp} \simeq  3.3$).
\item
In the remaining cases, we use our best estimate 
given by the results from Table~\ref{tableGM0}, corresponding
to the value of $G_M(0)$ which include the pion and kaon clouds 
(second column). 
\end{itemize}

For the discussion of the $q^2$-dependence of our model, 
we include also the model estimate of 
the Dalitz decay when we replace $G_M(q^2)$, 
by $\left. G_M(0) \right|_\pi$.
The results are represented by the dotted lines.
In the case of the $\Delta \to e^+ e^- N$ decay
we omit this estimate because it overlaps the 
estimate of the  constant form factor model (dashed line).

The importance of the pion cloud contributions 
is clearly shown in Fig.~\ref{figDGamma},
from the difference between the thick (total) 
and thin (bare) solid lines.
This difference of magnitude is a consequence 
of the relative magnitude of the corresponding estimates
for the transition form factors.
We recall that based on the estimates from Ref.~\cite{DecupletDecays},
also presented in Table~\ref{tablePionCloud},
the valence quark contributions to the transition form factors
are about 55\%--70\% of the total.
One concludes, then, that when the pion cloud contribution 
are about 50\% of the total, the bare estimates for $\frac{d \;}{d q}  \Gamma_{e^+e^-B}$
are about 1/4 of the total, since the decay widths 
are proportional to $|G_M|^2$.
This rough estimate is valid for most decays.
The main exceptions are the $\Sigma^{*+} \to e^+ e^- \Sigma^+$,
$\Sigma^{*0} \to e^+ e^- \Sigma^0$ and $\Xi^{*0} \to e^+ e^- \Xi^0$  decays, 
where the relative contribution of the core is 
larger (smaller pion cloud contributions).

The magnitudes of the different decays
can be clearly observed in the scale of the Dalitz decay widths:
large magnitudes for  
$\Delta \to e^+ e^- N$, $\Sigma^{\ast 0} \to e^+ e^- \Lambda$,
$\Sigma^{\ast +} \to e^+ e^- \Sigma^+$ and $\Xi^{\ast 0} \to e^+ e^- \Xi^0$
(scale $10^{-3}$);
intermediate magnitude for 
$\Sigma^{\ast 0} \to e^+ e^- \Sigma^0$ (scale $10^{-4}$);
small magnitudes for
$\Sigma^{\ast -} \to e^+ e^-  \Sigma^-$ and
$\Xi^{\ast -} \to e^+ e^-  \Xi^-$ 
(scale $10^{-5}$)~\cite{DecupletDecays}.
Those magnitudes are the consequence 
of the magnitudes of the magnetic form factors 
discussed in Sec.~\ref{secFormFactors}.

Concerning the comparison with the constant form factor 
model (dashed lines), 
one can conclude that the results are very close 
for the  $\Delta \to e^+ e^- N$ decay, for small values of $W$.
This happens because our model is compatible 
with the experimental value for $|G_M(0)|$,
as discussed earlier.
In the remaining cases, our result underestimates
the constant form factor model.
This underestimation is mainly a consequence of 
the non inclusion of the kaon cloud contribution 
in our $q^2$-dependent estimates,
in contrast with the constant form factor model.
This underestimation was discussed in 
detail in  Sec.~\ref{secFormFactors} for 
the $\Sigma^{* +} \to \gamma^\ast \Sigma^+$ form factor.

The impact of the form factor dependence on $q^2$
can be inferred from the comparison between 
the exact estimate (thick solid line) and the dotted line. 
As anticipated in Sec.~\ref{secFormFactors}, 
the $q^2$-dependence is more relevant for 
the $\Sigma^{* 0} \to e^+ e^- \Lambda$,
$\Sigma^{* -} \to e^+ e^- \Sigma^-$ 
and $\Xi^{* -} \to e^+ e^- \Xi^-$ decays.
The dominance of the exact result over the dotted line 
is clearly observed for large $q^2$, particularly for large values of $W$.

\begin{figure}[t]
\vspace{.3cm}
\centerline{
\mbox{
\includegraphics[width=3.0in]{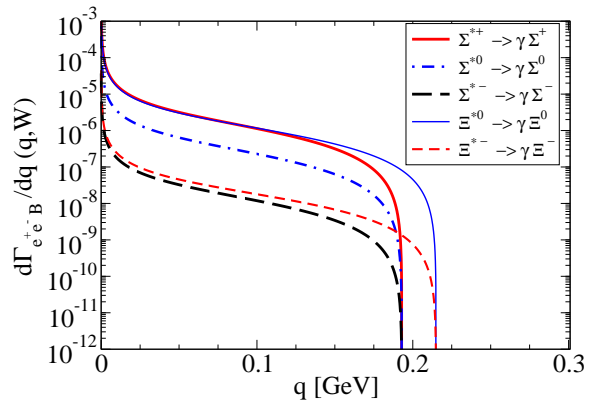} 
}}
\caption{\footnotesize{
Comparison between the $\Sigma^\ast$ and $\Xi^\ast$ 
Dalitz decay rates at the physical point 
($W \simeq 1.385$ GeV for $\Sigma^\ast$ 
and $W \simeq 1.533$ GeV for $\Xi^\ast$).}}
\label{figDGammaTotal}
\end{figure}

In Fig.~\ref{figDGammaTotal}, we compare 
the magnitudes of the $\Sigma^\ast$ and $\Xi^\ast$ Dalitz
decay rates at respective the mass poles.
Note the similarity between the 
results for $\Sigma^{\ast +} $/$\Xi^{\ast 0}$ decays, 
as well as  $\Sigma^{\ast -} $/$\Xi^{\ast -}$ decays 
in the region of $q^2$ where they can be compared.
These similarities are the consequence of the $SU(3)$ symmetry structure of 
the covariant spectator quark model,
combined with similar relative pion cloud contributions 
for the decays under discussion.
The relations between the valence quark contributions, $G_M^{\rm B}$,
given by  Eq.~(\ref{eqGMbare}) are explained by their 
dependence on the functions $j_i^S$,
which, according to Table~\ref{tableJS} are identical 
in the cases $\Sigma^{\ast +} $/$\Xi^{\ast 0}$ and $\Sigma^{\ast -} $/$\Xi^{\ast -}$.

The similarities between the $\Sigma^{\ast -} $ 
and $\Xi^{\ast -}$ decays are also explained by the 
$U$-spin symmetry~\cite{Keller12,DecupletDecays}, which is valid 
to the valence quark component of the transition form factors.
The $U$-spin symmetry, states that the decay transitions 
are similar when we replace a $d$-quark by a $s$-quark 
in the initial and final states~\cite{Keller12}.
The symmetry predicts also similar magnitudes
for the $\Delta \to \gamma \, N$ and  
the $\Sigma^{\ast 0} \to \gamma \, \Lambda$ 
Dalitz decay rates~\cite{Octet2Decuplet,DecupletDecays}.
This property, however, is not valid in the context of our model 
due to the difference of magnitudes of the pion cloud contributions
(larger in the first case).


\begin{figure*}[t]
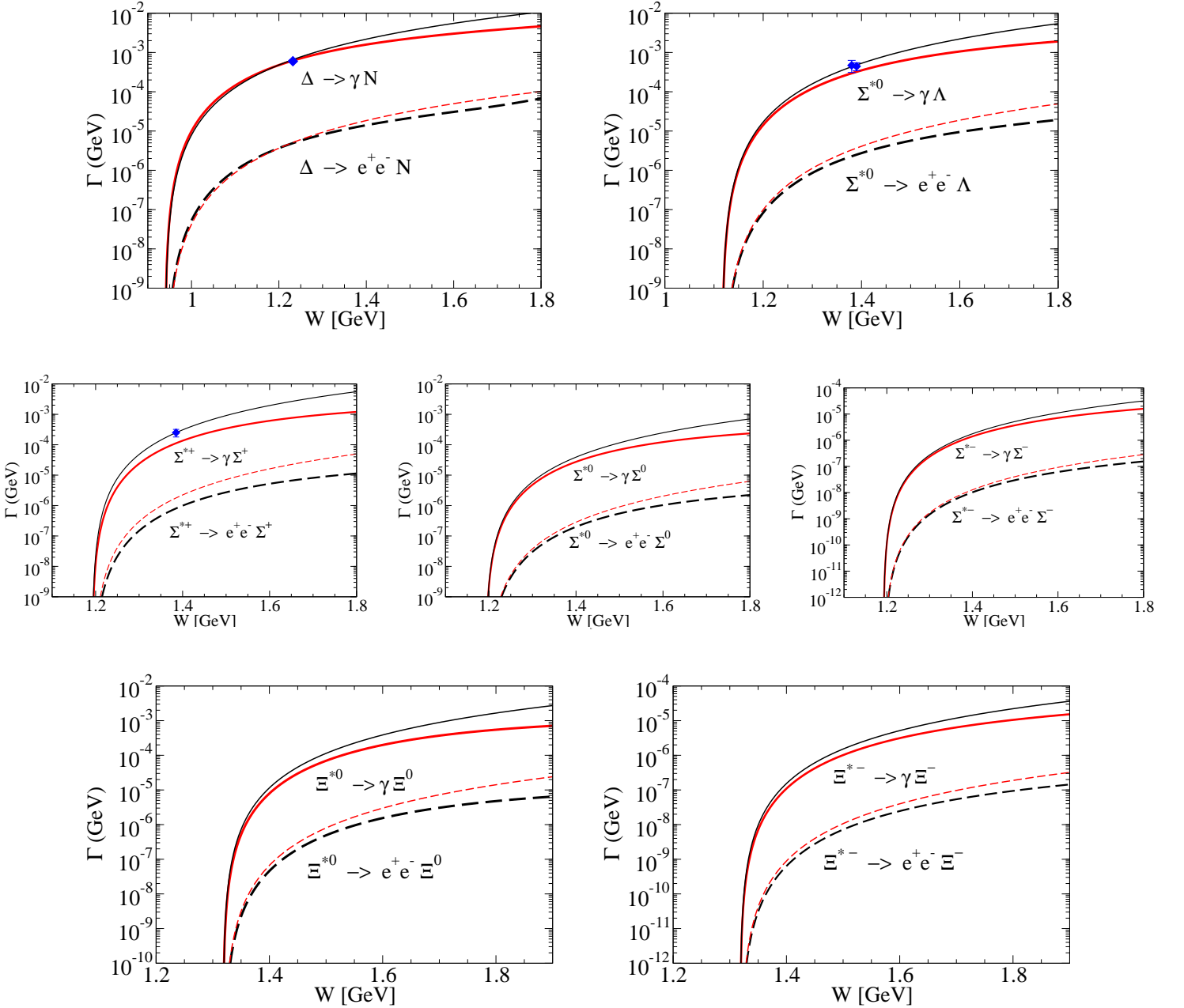

\centerline{\vspace{0.5cm}  }
\centerline{
\mbox{
\includegraphics[width=3.1in]{Gamma-Nucleon-v2} \hspace{.6cm}
\includegraphics[width=3.1in]{Gamma-Lambda-v2} }}
\vspace{.7cm}
\centerline{
\mbox{
\includegraphics[width=2.4in]{Gamma-SigmaP-v2} \hspace{.3cm}
\includegraphics[width=2.4in]{Gamma-Sigma0-v2} \hspace{.3cm}
\includegraphics[width=2.4in]{Gamma-SigmaM-v2} 
}}
\vspace{.7cm}
\centerline{
\mbox{
\includegraphics[width=3.1in]{Gamma-Xi0-v2} \hspace{.6cm}
\includegraphics[width=3.1in]{Gamma-XiM-v2} }}
\caption{\footnotesize{
Electromagnetic and Dalitz decay widths for all 
the decuplet decays in terms of $W$.
The thick lines represent our model.
The thin lines represent the constant form factor model.
Data from Table~\ref{tableGM0}.
}}
\label{figGamma}
\vspace{1.cm}  
\end{figure*}

Our model is compatible with the $U$-spin symmetry,
because it is based on an approximate $SU(3)$ flavor symmetry.
In the present case, the symmetry implies that the quark form factors 
associated with the $u$ quark (combination 
of isovector and isoscalar components) 
and the $s$ quark are similar at low $q^2$.
We recall, however, that the  $U$-spin symmetry 
is valid only for the valence quark component of the transition.
The covariant spectator quark model estimates
provide then a more consistent description 
of the radiative and Dalitz decays.

\subsection{Decay widths in terms of the invariant mass}
\label{secGammaW}

The results for the radiative ($B^\prime \to \gamma B$ )
and Dalitz ($B^\prime \to e^+ e^- B$) decay
widths in terms of $W$
are presented in Fig.~\ref{figGamma},
for all the decuplet baryon decays.
The thick lines represent our estimates.
The thin lines represent the estimates of 
the constant form factor model. 
We include also the data for $\Gamma_{\gamma B}$ 
at the physical mass in the cases:
$\Delta \to \gamma \, N$, $\Sigma^{\ast 0} \to \gamma \, \Lambda$ 
and  $\Sigma^{\ast +} \to \gamma \Sigma^+$, 
according to the results from Table~\ref{tableGM0}.

In the cases $\Sigma^{\ast 0} \to \gamma \, \Lambda$ 
and  $\Sigma^{\ast +} \to \gamma \, \Sigma^+$, one can notice  
some underestimation of the data.
This result is in part the consequence of 
including only the contribution of the pion cloud.
The inclusion of the kaon cloud approaches the model estimate 
to the data, as can be inferred also from Table~\ref{tableGM0}
(compare the second and the third columns).

We choose to not include the kaon cloud contributions 
on the radiative decays, because our extension for finite $q^2$ 
is justified, at the moment, only for the pion cloud contribution.
In general, the kaon cloud contributions are 
at most 20\% of the pion cloud contributions  
(Table~\ref{tableGM0}), except for $\Xi^{*0}$, 
where the effect of the kaon is about 25\% of the pion cloud.

Concerning the comparison with the results of the QED model,
there are two points to debate.
First, when we use a constant value for $|G_M|$, 
the results for $\Gamma_{\gamma B} (W)$ are close to 
the model estimates for  $\Gamma_{\gamma B}(W)$ for small values of $W$, 
and start to overestimate the model above 
a certain value of $W$.
Second, the overestimation of QED model, 
for large values of $W$ is expected due our model underestimation for $|G_M|$.
More definite conclusions can be drawn 
only when the unknown decay widths are determined experimentally.

Regarding the inclusion of $q^2$-dependent kaon cloud contributions,
our expectation is that the slope associated 
with the kaon cloud contributions is 
larger than the slope associated with
the pion cloud contributions near $q^2=0$,
since the kaon cloud effects are more suppressed 
than the pion cloud effects in the spacelike region\footnote{This 
effect can be better understood 
assuming that the meson cloud contributions to 
the transition form factors can, near $q^2=0$, 
be simulated by a multipole function 
$1/(1 + Q^2/\Lambda^2)^n$, where $n> 2$ is an integer
(the exact value is not important for the discussion) 
and $\Lambda^2$ is a cutoff of $Q^2$.
One conclude, then, that the cutoff associated with 
the kaon cloud is smaller than the one associated to the pion, 
since the kaon cloud effects are suppressed more strongly
than the pion cloud effects.
The consequence of this relation between cutoffs 
is that the magnitude of the derivative of the kaon cloud multipole
($\propto 1/\Lambda^2$) at $q^2=0$ is larger than 
the magnitude of the derivative of the pion cloud multipole.}.
The consequence of this trend is that the
transition form factors are expected to be enhanced in the 
timelike region 
with the inclusion of the kaon cloud contributions.
We recall, however, that the kaon cloud contributions near $q^2=0$ 
are at most about 20\% of the pion cloud contributions.
There is then the possibility that the kaon cloud 
effects may not be very relevant in the region $0 < q^2 \le (W -M_B)^2$.
Only more detailed calculations can determine how important 
may be the enhancement of the transition form factors 
due to the kaon cloud effects, in the timelike region.

Our estimates for the $B' \to e^+ e^- B$ Dalitz decays 
underestimate, in general, the QED model.
This tendency is a consequence of the results obtained 
for the Dalitz decay rates (Fig.~\ref{figDGamma}),
where the QED model overestimates, in general, 
the covariant spectator quark model.
The exception is the $\Delta \to e^+ e^- N$ decay,
where our model and the QED estimates are close.

It is worth noticing, that only the $\Delta(1232)$ Dalitz decay 
was measured experimentally at the pole ($W \simeq 1.232$ GeV).
Our estimate of the $\Delta(1232)$ Dalitz decay width 
is consistent with the result of HADES~\cite{HADES17}. 
All estimates for the $B' \to e^+ e^- B$ Dalitz decays at 
the physical decuplet baryon mass ($M_{B'}$) are 
presented in Table~\ref{tableDelitz}.
Excluding the $\Delta(1232)$, the remaining estimates 
are predictions to be tested by future experiments.

The $\Sigma^{*0} \to  e^+ e^- \Lambda$ decay width was also  
estimated within the chiral perturbation theory combined with 
dispersion relations~\cite{Junker20a},
obtaining a slightly larger value (3.0--3.4 keV).
The inclusion of the kaon cloud effects in our framework 
can also increase our estimate.

\begin{table}[t]
\begin{tabular}{l  c }
\hline
\hline
 Decay   &   $\Gamma_{e^+ e^- B}$ (keV)      \\
\hline
\hline
$\Delta \to e^+ e^- N$      &   4.9  \\[.25cm]
$\Sigma^{*0} \to  e^+ e^- \Lambda$   &   2.4  \\[.25cm]
$\Sigma^{*+} \to  e^+ e^- \Sigma^+$   &   0.81 \\
$\Sigma^{*0} \to  e^+ e^- \Sigma^0$   &   0.16 \\
$\Sigma^{*-} \to  e^+ e^- \Sigma^-$   &   0.83$\times 10^{-3}$ \\[.25cm]
$\Xi^{*0} \to  e^+ e^- \Xi^0$   &   0.76 \\
$\Xi^{*-} \to  e^+ e^- \Xi^-$   &    1.2$\times 10^{-3}$ \\[.1cm]
\hline
\hline
\end{tabular}
\caption{\footnotesize
Decuplet baryon Dalitz decay widths. 
The HADES result for the $\Delta(1232)$ Dalitz decay 
is $4.90 \pm 0.83$ keV~\cite{HADES17}.}
\label{tableDelitz}
\end{table}

\section{Outlook and conclusions}
\label{secConclusions}

The HADES facility provides a rare opportunity 
to study electromagnetic transitions between baryon 
states in the timelike region ($q^2 > 0$). 
Those experiments complement the 
information obtained from electro-production of baryon resonances 
in the spacelike region ($q^2 \le 0$).
The recent and the upcoming results from HADES motivate 
the development of theoretical models 
for the $\gamma^\ast B \to B'$ transition form factors
in the timelike region, where $B$ and $B'$ are generic baryons.

Of particular interest are the Dalitz decays 
of baryons ($B'\to e^+ e^- B$), including hyperons.
Measurements of the $\Delta(1232)$ Dalitz decays have been reported recently.
The analysis of the $\Sigma^0(1385) \to e^+ e^-\Lambda(1116)$ decay
is expected in a near future.
Due to the capability of HADES to produce hyperons,
other decuplet baryon Dalitz decays are expected 
to be measured in the following years.
The next natural candidate, based on the 
estimated magnitude, is the 
$\Sigma^+(1385) \to e^+ e^- \Sigma^+(1193)$ decay.

To complement the experimental activity at HADES,
we present here model estimates for the Dalitz decay rates 
and Dalitz decay widths for all decuplet baryons.
Our calculations are based on the covariant
spectator quark model for the 
octet baryon to decuplet baryon electromagnetic transitions,
extended in the present work to the timelike region.
The model was previously calibrated by lattice QCD data
for the baryon octet and baryon decuplet,
and takes into account the 
pion cloud dressing of the baryon cores.
The model is successful in the description 
of the radiative decays:
$\Delta(1232) \to \gamma N$, 
$\Sigma^0(1385) \to \gamma \, \Lambda (1116)$ 
and $\Sigma^+(1385) \to \gamma \,  \Sigma^+ (1193)$.
Under study is the extension of the present model 
with the inclusion of the kaon cloud contribution for finite $q^2$,
which may approach the model estimates to the data.

We conclude that, in general, the valence quark effects
give the dominant contribution to the transition form factors 
and to the Dalitz decay widths,
but that pion cloud contribution provides  
significant corrections, which improve the description of the data.
In most cases, the pion cloud effects contribute 
with about 30\%--45\% to the transition form factors near $q^2=0$.
In some cases, those contributions are only about 20\%
($\Sigma^+(1385)$,  $\Sigma^0(1385)$  and $\Xi^0 (1530)$ decays).

We conclude also, that different magnitudes are 
expected to the radiative and Dalitz decay widths 
according with valence quark content:
large magnitudes for the $\Delta(1232)$, 
$\Sigma^0(1385) \to e^+ e^- \Lambda(1116)$, $\Sigma^+(1385)$ and  $\Xi^0(1530)$ decays; 
intermediate magnitudes for the  
$\Sigma^0(1385) \to e^+ e^- \Sigma^0(1193)$ decay;
small magnitudes for the $\Sigma^-(1385)$ and $\Xi^-(1530)$ decays.
We observed also that the $\Sigma^+(1385)$ and $\Xi^0(1530)$ decays,
as well as the  $\Sigma^-(1385)$ and $\Xi^-(1530)$ decays,
have similar Dalitz decay rates.

We also analyze the role of the $q^2$-dependence 
of the form factors.
We conclude that, in general, the QED approach 
(constant form factor model) is not a good approximation, 
as already observed in the case of the $\Delta(1232)$ Dalitz decay. 
The impact of  the $q^2$-dependence of the form factors is, however, 
less significant than in the case of the $\Delta(1232)$.
The $\Sigma^0(1385) \to \gamma^\ast \Lambda(1116)$ transition form factors 
are enhanced in the timelike region due to the pion cloud effects.
The $q^2$-dependence is also relevant for the  
$\Sigma^- (1385) \to e^+ e^- \Sigma^- (1193)$ and  
$\Xi^- (1530) \to e^+ e^- \Xi^- (1318)$ decays.

The covariant spectator quark model 
proved also to be a useful framework to 
study Dalitz decays of nucleon excited states ($N^\ast$),
more specifically in the cases of the 
$\Delta(1232)$, $N(1520)$ and $N(1535)$ 
resonances~\cite{Timelike2,N1520TL,N1535TL}.
Under study is the possibility of extending 
the formalism to other baryon systems, 
which may also be regarded as a combination of 
valence quark cores combined with meson cloud excitations 
of the baryon cores. 

\begin{acknowledgments}
G.R.~thanks Piotr Salabura for helpful discussions and suggestions.
G.R.~was supported by the Funda\c{c}\~ao de Amparo \`a
Pesquisa do Estado de S\~ao Paulo (FAPESP):
Project No.~2017/02684-5, Grant No.~2017/17020-BCO-JP.
\end{acknowledgments}

\appendix

\section{Quark form factors}
\label{appQuarkFF}

Motivated by the VMD mechanism, we use the following 
parametrizations for the quark form factors 
$f_{i0}$ and $f_{i\pm}$ ($i=1,2$)
\ba
& &
\hspace{-1.2cm}
f_{1-}(q^2)  = \lambda_q + (1 - \lambda_q) \frac{m_\rho^2}{m_\rho^2 - q^2} 
- c_- \frac{M_h^2 q^2}{(M_h^2 - q^2)^2} 
\label{eqF1m}\\
&&
\hspace{-1.2cm}
f_{1+}(q^2) = \lambda_q + (1 - \lambda_q) \frac{m_\omega^2}{m_\omega^2 - q^2} 
- c_+ \frac{M_h^2 q^2}{(M_h^2 - q^2)^2} 
\label{eqF1p}\\
& &
\hspace{-1.2cm}
f_{10} (q^2) = \lambda_q + (1 - \lambda_q) \frac{m_\phi^2}{m_\phi^2 - q^2} 
- c_0 \frac{M_h^2 q^2}{(M_h^2 - q^2)^2} 
\label{eqF10}\\
&&
\hspace{-1.2cm}
f_{2-}(q^2) =   \kappa_- \left\{ 
d_- 
 \frac{m_\rho^2}{m_\rho^2 - q^2} 
+ (1- d_-)  \frac{M_h^2}{M_h^2 - q^2}  \right\} 
\label{eqF2m}\\
&&
\hspace{-1.2cm}
f_{2+}(q^2) =   \kappa_+ \left\{ 
d_+ 
 \frac{m_\omega^2}{m_\omega^2 - q^2} 
+ (1- d_+)  \frac{M_h^2}{M_h^2 - q^2}  \right\} \\
&&
\hspace{-1.2cm}
f_{20}(q^2) =   \kappa_0 \left\{ 
d_0 
 \frac{m_\phi^2}{m_\phi^2 - q^2} 
+ (1- d_0)  \frac{M_h^2}{M_h^2 - q^2}  \right\}, 
\label{eqF20}
\ea 
where $m_\rho$, $m_\omega$ and $m_\phi$ 
represent the masses of the mesons 
$\rho$, $\omega$ and $\phi$, respectively.
The terms with $M_h$ correspond to an effective heavy vector meson
which parametrize the short range effects.
The value of $M_h$ is fixed as $M_h = 2 M_N$~\cite{Nucleon,Lattice}.
In numerical calculations, we use the 
approximation $m_\omega = m_\rho$ for simplicity.

In Eqs.~(\ref{eqF2m})--(\ref{eqF20}), 
$\kappa_q$ represent quark anomalous magnetic moments.
We use the parametrization derived from the study 
of the octet and decuplet baryons~\cite{Octet2,Omega}.
We take in particular $\kappa_- = 1.435$,  
$\kappa_+ = 1.803$ and $\kappa_0 = 1.462$.
To convert to the flavors  $q=u,d,s$, 
one uses $\kappa_u = \frac{1}{4}(\kappa_+ + 3 \kappa_-)$,
$\kappa_d = \frac{1}{2}(2 \kappa_- - \kappa_+)$ 
and $\kappa_s= \kappa_0$~\cite{Nucleon,Omega}.

In the equations $\lambda_q$ is a parameter related with 
the quark density number in deep inelastic scattering~\cite{Nucleon}.
The numerical value is $\lambda_q =1.21$.
The remaining parameters are 
$c_+= 4.160$, $c_-=1.160$, $c_0= 4.427$,
$d_+= d_- =-0.686$ and $d_0=-1.860$~\cite{Octet2,Omega}.

The expressions (\ref{eqF1m})--(\ref{eqF20}) are valid 
in the region $q^2 < 0$, when the
vector meson decay widths vanish, $\Gamma_v  \equiv 0$ ($v=\rho, \omega,\phi$).
For the extension of the quark form factors 
to the timelike region ($q^2 > 0$), we consider 
the replacement ($v=\rho, \omega, \phi$)
\ba
\frac{m_v^2}{m_v^2 -q^2} 
\to 
 \frac{m_v^2}{m_v^2 -q^2 - i m_v \Gamma_v (q^2)}. 
\ea
The decay width functions $\Gamma_v(q^2)$, which describe 
the dressing of the vector mesons in terms 
of the possible meson decay channels, are discussed next.

Following our previous works based on 
the $\Delta(1232)$ Dalitz decay, 
we consider for the isovector components 
($\rho$-pole) the function~\cite{Timelike,Timelike2,Muhlich,Donges95}
\ba
\Gamma_\rho(q^2)=
\Gamma_\rho^0 \frac{m_\rho^2}{q^2} 
\left(\frac{q^2 - 4 m_\pi^2}{m_\rho^2- 4 m_\pi^2} \right)^{3/2}
\theta(q^2 - 4 m_\pi^2),
\label{eqGammaRho}
\ea 
where $\Gamma_\rho^0 = 0.149$ GeV.
The previous equation parametrize 
the width associated to the decay $\rho \to 2 \pi$
for a virtual $\rho$ with square four-momentum $q^2$~\cite{Connell97,Muhlich}.
Alternative parametrizations for $\Gamma_\rho (q^2)$ are 
presented in Refs.~\cite{Weil12,GammaR1,GammaR3,GammaR4}.

For the isoscalar channel, associated with the $\omega$-meson,
one needs to consider the combination 
of the decays $\omega \to 2\pi$ and $\omega \to 3 \pi$.
Following our work on the $N(1520)$ Dalitz decay~\cite{N1520TL},
we decompose  
\ba
\Gamma_\omega (q^2) = 
\Gamma_{2 \pi} (q^2) + \Gamma_{3 \pi} (q^2),
\ea
where the first term parametrize the decay  $\omega \to 2\pi$ a
and the second term  parametrize the decay  $\omega \to 3\pi$.
The expression for $\Gamma_{2 \pi} (q^2) $ is similar 
to $\Gamma_{\rho} (q^2) $ except for the strength~\cite{N1520TL,Muhlich}.
As for the decay  $\omega \to 3\pi$, 
we consider a model based on the  
process $\omega \to \rho \pi \to 3 \pi$,
where the intermediate $\rho$ decays into 2 pions~\cite{Muhlich}.
We do not reproduce here the expressions for $\Gamma_{2 \pi}$ 
and $\Gamma_{3 \pi}$, since they can be found in Ref.~\cite{N1520TL}. 
We just point out that the $3 \pi$ channel dominates 
for $q^2 > 0.55$ GeV$^2$.
A more detailed discussion of $\Gamma_\omega (q^2)$
is presented in  Ref.~\cite{N1520TL}.

Finally, for the $\phi$ decay width, we consider 
the simplified  parametrization
\ba
\Gamma_\phi(q^2) = \Gamma_\phi^0 
\frac{m_\phi^2}{q^2} 
\left(\frac{q^2 - 4 m_K^2}{m_\phi^2- 4 m_K^2} \right)^{3/2}
\theta(q^2 - 4 m_K^2), \nonumber\\
\label{eqGammaPhi}
\ea
where $\Gamma_\phi^0 = 4.23 \times 10^{-3}$ GeV, 
and $m_K$ is the kaon mass ($m_K \simeq 0.5$ GeV).
Equation (\ref{eqGammaPhi}) describes the $\phi \to 2 K$ 
($K$ is the kaon) under the assumption that it is the dominate 
decay of the $\phi$. 
According with PDG the $2K$ decays correspond to 
about 85\% of the $\phi$ decays~\cite{PDG18}.

For the range of the calculation  of the present 
work ($W < 2$ GeV)
the regularization of the $\phi$ pole is not 
very relevant, since $m_\phi^2 \simeq 1$ GeV$^2$ $\gg q^2$.
The singularities associated to the 
 $\phi$-meson appear, then only for $W \ge M_B + m_\phi > 2.1$ GeV.
Nevertheless, we regularize the $\phi$-propagator for consistence.
We note also that even the calculations 
more dependent on the $\phi$-pole, in particular 
the Dalitz decay widths $\Gamma_{e^+ e^- B} (W)$, 
are weakly dependent on the shape of $\Gamma_\phi(q^2)$.

We also concluded that the Dalitz decay widths $\Gamma_{e^+ e^- B} (W)$
depend weakly of the explicit form used for $\Gamma_\rho (q^2)$ 
in Eq.~(\ref{eqGammaRho}). 
Equivalent results can be obtained 
when we replace $\frac{m_\rho^2}{q^2}$ 
by $\frac{m_\rho}{q}$, following Refs.~\cite{GammaR1}.
The main differences appear only near $q^2=m_\rho^2$, 
and their effects are diluted 
in the integration in $q$.

\section{Regularization of high mass poles}
\label{appRegularization}

For a given $W$ the square momentum $q^2$ is limited 
by the kinematic condition $q^2 \le (W-M_B)^2$.
If there is a singularity at $q^2=\Lambda^2$ 
the singularity will appear for values of $W$ 
such that $(W- M_B)^2 \ge q^2$, or
$W  \ge M_B + \Lambda$.

To avoid those singularities, for single poles 
with a generic momentum scale $\Lambda$,
we use the following procedure
\ba
\frac{\Lambda^2}{\Lambda^2 -q^2} \to 
\frac{\Lambda^2}{\Lambda^2 -q^2 - i \Lambda \Gamma_X(q^2)},
\ea  
where 
\ba
\Gamma_X(q^2)= 4 \Gamma_X^0
\left( \frac{q^2}{q^2+ \Lambda^2}
\right)^2
\theta(q^2), 
\label{eqGammaX}
\ea
In the last equation $\Gamma_X^0$ is a constant 
given by  $\Gamma_X^0 = 4 \Gamma_\rho^0\simeq 0.6$ GeV.

This procedure is used on the pole $q^2=M_h^2$ 
of the quarks form factors, for the single pole 
(Pauli form factors) and double pole (Dirac form factors).

For powers of monopole factors used in the pion cloud contribution (\ref{eqGMpi}), 
we approximate the result by the magnitude of the expression:
\ba
\left( \frac{\Lambda^2}{\Lambda^2 -q^2}
\right)^n 
\to 
\left( \frac{\Lambda^4}{(\Lambda^2 -q^2)^2 + \Lambda^2 [\Gamma_X(q^2)]^2}
\right)^{\frac{n}{2}}, 
\ea
where $\Gamma_X(q^2)$ is determined by Eq.~(\ref{eqGammaX}).

In the generalization of the dipole function $\tilde G_D$
defined by Eq.~(\ref{eqGD}), 
we use Eq.~(\ref{eqGammaX}), with $\Lambda= \Lambda_D$.

\setcounter{table}{0}
\renewcommand{\thetable}{C\arabic{table}}

\section{Calculation of the meson cloud contributions}
\label{appMC}

We present here a brief revision of the 
calculation of the contributions of the diagrams (a) and (b) 
from Fig.\ref{figMesonCloud}, following Ref.~\cite{DecupletDecays}.

The calculations of the meson cloud contributions 
are based on the cloudy bag model (CBM).
Since those contributions depend on the photon couplings 
with the bare baryons, it is necessary to make the 
connection between the Dirac and Pauli couplings between CBM and the 
covariant spectator quark model.
This connection was performed in Ref.~\cite{DecupletDecays}
with the comparison of the results from both 
frameworks for the octet baryon to decuplet baryon transitions.
One obtains the same result 
for the magnetic transition form factor at low $Q^2$
in both frameworks, when we define the quark ($q=u,d,s$) 
effective magnetic moments as
\ba
\mu_q = 
\sqrt{\frac{2}{3}} 
\left[
\frac{2 M_B}{M_{B'} + M_B} + 
\frac{M_B}{M_N} \kappa_q
\right] {\cal I}(0,M_{B'}),
\label{eqMUQ}
\ea 
where ${\cal I}(0, M_{B'})$ 
is defined by Eq.~(\ref{eqIntBBp}).
Notice that the value of $\mu_q$ depends on the explicit transition.
In the static limit, where all baryons are very heavy 
and the mass differences can be neglected, 
one obtains $\mu_q \propto (1 + \kappa_q)$.

In Eq.~(\ref{eqMUQ}), the presence of the 
overlap integral integral is important because 
it tend to reduce the contribution of the bare core
when we use different radial wave functions for 
the octet baryon and decuplet baryon.
In an exact $SU(3)$ model where octet and decuplet 
baryons are described by the same radial wave functions 
(also no mass difference), we obtain  
${\cal I}(0, M_{B'}) =1$~\cite{DecupletDecays}.

We can now describe the calculations 
of the meson cloud contributions from the diagrams 
Fig.\ref{figMesonCloud}(a) and (b) to the 
magnetic form factors.

\begin{table}[t]
\begin{center}
\begin{tabular}{l  c   c}
\hline
\hline 
         &   $\pi$    &  $K$  \\
\hline 
$\Delta \to \gamma^\ast N$ & $N $, $\Delta $ & 
                            $\Sigma$, $\Sigma^\ast$ \\  
$\Sigma^{\ast 0}  \to \gamma^\ast \Lambda $ &
$\Sigma$, $\Sigma^\ast $ &
$N$, $\Xi$,  $\Xi^\ast$ \\
$\Sigma^{\ast }  \to \gamma^\ast \Sigma $ \spQ & 
\spQ
$\Lambda$, $\Sigma$, $\Sigma^\ast$ \spQ &
\spQ $N$, $\Delta$,   $\Xi$,  $\Xi^\ast$ \spQ\\
$\Xi^\ast \to \gamma^\ast \Xi$ &
$\Xi $, $\Xi^\ast$ &
\spQ 
$\Lambda$, $\Sigma$,  
$\Sigma^\ast$,  $\Omega$ \spQ\\
\hline
\hline
\end{tabular}
\end{center}
\caption{\footnotesize
Diagram a: $B_1$ contributions 
for the $B^\prime \to \gamma^\ast B$ decay.
There are contributions for $M=\pi$ and $M=K$.}
\label{tab-diagramA}
\end{table}

\subsection{Diagram (a)}

\begin{table*}[t]
\begin{center}
\begin{tabular}{l  c   c  c}
\hline
\hline 
         &   $\pi$    &  $K$  &   $\eta$\\
\hline 
$\Delta \to \gamma^\ast N$ & $N N $, $N \Delta$ & 
      $\Lambda \, \Sigma$, $\Lambda \, \Sigma^\ast$, $\Sigma \, \Sigma $  &  $N \Delta $  \\  
               & $\Delta N$, $\Delta \Delta $  & 
  $\Sigma \,  \Sigma^\ast $,   $\Sigma^\ast  \Sigma$,  $\Sigma^\ast \Sigma^\ast$  & 
\\[.2cm]
$\Sigma^{\ast 0}  \to \gamma^\ast \Lambda $ &
$\Sigma \, \Lambda$, $\Sigma  \,\Sigma $, $ \Sigma \, \Sigma^\ast $           &  
 $N N$,   $ N \Delta $ &  $\Lambda \Sigma$\\
& 
\spQ $\Sigma^\ast \Lambda$,  $\Sigma^\ast \Sigma$, $\Sigma^\ast \Sigma^\ast$ \spQ &
 $\Xi \, \Xi$,  $\Xi \, \Xi^\ast $,  $\Xi^\ast  \Xi$, $\Xi^\ast \Xi^\ast $
                   &   \\ [.2cm]
$\Sigma^{\ast }  \to \gamma^\ast \Sigma $ & 
$\Sigma \, \Lambda $, $\Sigma \, \Sigma$, $\Sigma \, \Sigma^\ast$ &
 \spQ $N N$, $N \Delta$, $\Delta N$, $\Delta \Delta$ \spQ 
  &  $\Sigma \, \Sigma$,  $\Sigma^\ast \Sigma$\\
 & 
$\Lambda \, \Lambda$, $\Lambda \, \Sigma$, $\Lambda \, \Sigma^\ast$&  
\spQ $\Xi \, \Xi$,  $\Xi \, \Xi^\ast $,  $\Xi^\ast  \Xi$, $\Xi^\ast \Xi^\ast $ \spQ
 \\
 & 
$\Sigma^\ast \Lambda$, $\Sigma^\ast  \Sigma$,  $\Sigma^\ast \Sigma^\ast$ & \\[.2cm]
$\Xi^\ast \to \gamma^\ast \Xi$ & $\Xi \, \Xi$, $\Xi \, \Xi^\ast$   
 & $\Sigma \, \Sigma$, $\Sigma \, \Sigma^\ast$, $\Sigma^\ast \Sigma$, $\Sigma^\ast \Sigma^\ast$
 & $\Xi \, \Xi$, $\Xi \, \Xi^\ast$  \\
& $\Xi^\ast \Xi$, $\Xi^\ast \Xi^\ast$  
 & $\Lambda  \, \Lambda$, $\Lambda \, \Sigma$, $\Lambda \, \Sigma^\ast$  
&  $\Xi^\ast \Xi$, $\Xi^\ast \Xi^\ast$  \\
& & 
$\Sigma \, \Lambda$, $\Sigma^\ast \Lambda$, $\Omega \, \Omega$  
 & \\
\hline
\hline
\end{tabular}
\end{center}
\caption{\footnotesize
Diagram (b): contributions $B^\prime B_1 B_2 B$ 
for the $B^\prime \to \gamma^\ast B$ decay.
There are contributions for $M=\pi$, $M=K$ and $M= \eta$.}
\label{tab-diagramB}
\end{table*}

The calculation of the  contributions 
for the diagram \ref{figMesonCloud}(a) 
are performed based on
\ba
G_M^{{\rm MC}a} =
\sum_{M,B_1} C_{B B';B_1}^M H_{B B'}^M(B_1),
\ea 
where $M$ labels the intermediate meson states 
($M=\pi, K$),
$C_{B B';B_1}^M$ are coefficients in the CBM framework, 
and $H_{BB'}^M(B_1)$ is the CBM loop integral 
associated to a diagram with an intermediate baryon $B_1$
and the meson $M$.
The function $H_{B B'}^M(B_1)$ is defined by 
Eq.~(4.2) from Ref.~\cite{DecupletDecays}.

The labels of the state $B_1$ used 
in the calculations are displayed in Table~\ref{tab-diagramA}.
The couplings associated to the states are presented 
in Table IV from Ref.~\cite{DecupletDecays}.
For completeness, we present also the intermediate states 
associated to the kaon.

\subsection{Diagram (b)}

The calculations of the contributions 
for the diagram \ref{figMesonCloud}(b)
are performed based on
\ba
G_M^{{\rm MC}b} =
\sum_{M,B_1,B_2} D_{B B';B_1,B_2}^M H^{2M}_{B B'}(B_1,B_2),
\ea 
where $M$ labels the intermediate meson states 
($M=\pi, K, \eta$), $D_{B B';B_1,B_2}^M$ are coefficients 
in the CBM framework,
and $H^{2M}_{B B'}(B_1,B_2)$ is the CBM loop integral 
associated to a diagram with the intermediate baryons $B_1$, $B_2$
and the meson $M$.
The integral $H^{2M}_{B B'}(B_1,B_2)$ is defined by Eq.~(4.4) 
in Ref.~\cite{DecupletDecays}.

The function $G_M^{{\rm MC}b}$ include the contributions
of the baryons $B_1$, $B_2$ displayed in Table~\ref{tab-diagramB}.
The explicit expressions for $D_{B B';B_1,B_2}^M$
are linear combinations of the 
effective quark form factors $\mu_q$  and are presented in 
the Appendix A of Ref.~\cite{DecupletDecays}.

The dependence of the diagram \ref{figMesonCloud}(b)
contributions on the intermediate bare states are 
then expressed by the dependence on the effective quark form factors.

Note that in intermediate state, one has all kinds 
of baryon transitions: octet to octet, 
octet to decuplet, decuplet to octet 
and decuplet to decuplet.



\begin{thebibliography}{00}



\bibitem{NSTAR} 
  I.~G.~Aznauryan {\it et al.},
  Int.\ J.\ Mod.\ Phys.\ E {\bf 22}, 1330015 (2013)
  [arXiv:1212.4891 [nucl-th]].


\bibitem{Aznauryan12} 
  I.~G.~Aznauryan and V.~D.~Burkert,
  Prog.\ Part.\ Nucl.\ Phys.\  {\bf 67}, 1 (2012)
  [arXiv:1109.1720 [hep-ph]].


\bibitem{NSTAR17} 
  G.~Ramalho,
  Few Body Syst.\  {\bf 59}, 92 (2018)
  [arXiv:1801.01476 [hep-ph]].



\bibitem{Burkert04} 
  V.~D.~Burkert and T.~S.~H.~Lee,
  Int.\ J.\ Mod.\ Phys.\ E {\bf 13}, 1035 (2004)
  [nucl-ex/0407020].



\bibitem{Drechsel07} 
  D.~Drechsel, S.~S.~Kamalov and L.~Tiator,
  Eur.\ Phys.\ J.\ A {\bf 34}, 69 (2007)
  [arXiv:0710.0306 [nucl-th]].



\bibitem{PDG18} 
  M.~Tanabashi {\it et al.} [Particle Data Group],
  Phys.\ Rev.\ D {\bf 98}, 030001 (2018).



\bibitem{Hyperons} 
  G.~Ramalho, M.~T.~Pe\~na and K.~Tsushima,
  Phys.\ Rev.\ D {\bf 101}, 014014 (2020)
  [arXiv:1908.04864 [hep-ph]].



\bibitem{Octet4} 
  F.~Gross, G.~Ramalho and K.~Tsushima,
  Phys.\ Lett.\ B {\bf 690}, 183 (2010)
  [arXiv:0910.2171 [hep-ph]].




\bibitem{Octet1} 
  G.~Ramalho and K.~Tsushima,
  Phys.\ Rev.\ D {\bf 84}, 054014 (2011)
  [arXiv:1107.1791 [hep-ph]].


\bibitem{Octet2} 
  G.~Ramalho, K.~Tsushima and A.~W.~Thomas,
  J.\ Phys.\ G {\bf 40}, 015102 (2013)
  [arXiv:1206.2207 [hep-ph]];
  G.~Ramalho, J.~P.~B.~C.~de Melo and K.~Tsushima,
  Phys.\ Rev.\ D {\bf 100}, 014030 (2019)
  [arXiv:1902.08844 [hep-ph]].




\bibitem{LambdaSigma0} 
  G.~Ramalho and K.~Tsushima,
  Phys.\ Rev.\ D {\bf 86}, 114030 (2012)
  [arXiv:1210.7465 [hep-ph]].


  
\bibitem{Octet2Decuplet} 
  G.~Ramalho and K.~Tsushima,
  Phys.\ Rev.\ D {\bf 87}, 093011 (2013)
  [arXiv:1302.6889 [hep-ph]].


\bibitem{DecupletDecays} 
  G.~Ramalho and K.~Tsushima,
  Phys.\ Rev.\ D {\bf 88}, 053002 (2013)
  [arXiv:1307.6840 [hep-ph]].







\bibitem{HADES14} 
  G.~Agakishiev {\it et al.},
  Eur.\ Phys.\ J.\ A {\bf 50}, 82 (2014)
  [arXiv:1403.3054 [nucl-ex]].



\bibitem{HADES17} 
  J.~Adamczewski-Musch {\it et al.} [HADES Collaboration],
  Phys.\ Rev.\ C {\bf 95}, 065205 (2017)
  [arXiv:1703.07840 [nucl-ex]].




\bibitem{HADES17b} 
  J.~Adamczewski-Musch {\it et al.} [HADES Collaboration],
  Eur.\ Phys.\ J.\ A {\bf 53}, 149 (2017)
  [arXiv:1703.08575 [nucl-ex]].




\bibitem{Salabura13} 
  P.~Salabura {\it et al.} [HADES Collaboration],
  J.\ Phys.\ Conf.\ Ser.\  {\bf 420}, 012013 (2013).




\bibitem{ColeTL} 
  P.~Cole, B.~Ramstein and A.~Sarantsev,
  Few Body Syst.\  {\bf 59}, 144 (2018).



\bibitem{Weil12} 
  J.~Weil, H.~van Hees and U.~Mosel,
  Eur.\ Phys.\ J.\ A {\bf 48}, 111 (2012)
  Erratum: [Eur.\ Phys.\ J.\ A {\bf 48}, 150 (2012)]
  [arXiv:1203.3557 [nucl-th]].


\bibitem{Ramstein18a} 
  B.~Ramstein [HADES Collaboration],
  Few Body Syst.\  {\bf 59}, 141 (2018).




\bibitem{HADES20c} 
  J.~Adamczewski-Musch {\it et al.} [HADES Collaboration],
  Phys.\ Rev.\ C {\bf 102}, 024001 (2020)
  [arXiv:2004.08265 [nucl-ex]].


  
\bibitem{Salabura20} 
  P.~Salabura and J.~Stroth,
  arXiv:2005.14589 [nucl-ex].






\bibitem{Ramstein18b} 
  B.~Ramstein,
  Few Body Syst.\  {\bf 59}, 143 (2018).


\bibitem{HADES17c} 
  J.~Adamczewski-Musch {\it et al.} [HADES Collaboration],
  Eur.\ Phys.\ J.\ A {\bf 53}, 188 (2017).




\bibitem{Ramstein19} 
  B.~Ramstein {\it et al.} [HADES Collaboration],
  EPJ Web Conf.\  {\bf 199}, 01008 (2019).



\bibitem{Lutz03} 
  M.~F.~M.~Lutz, B.~Friman and M.~Soyeur,
  Nucl.\ Phys.\ A {\bf 713}, 97 (2003)
  [nucl-th/0202049].



\bibitem{Shyam10} 
  R.~Shyam and U.~Mosel,
  Phys.\ Rev.\ C {\bf 82}, 062201(R) (2010)
  [arXiv:1006.3873 [hep-ph]].


\bibitem{Zetenyi12} 
  M.~Zetenyi and G.~Wolf,
  Phys.\ Rev.\ C {\bf 86}, 065209 (2012)
  [arXiv:1208.5671 [nucl-th]].













\bibitem{Dohrmann10}
  F.~Dohrmann {\it et al.},
  Eur.\ Phys.\ J.\  A {\bf 45}, 401 (2010)
  [arXiv:0909.5373 [nucl-ex]].




\bibitem{Timelike2} 
  G.~Ramalho, M.~T.~Pe\~na, J.~Weil, H.~van Hees and U.~Mosel,
  Phys.\ Rev.\ D {\bf 93}, 033004 (2016)
  [arXiv:1512.03764 [hep-ph]].







\bibitem{N1520TL} 
  G.~Ramalho and M.~T.~Pe\~na,
  Phys.\ Rev.\ D {\bf 95}, 014003 (2017)
  [arXiv:1610.08788 [nucl-th]].


\bibitem{N1535TL}
  G.~Ramalho and M.~T.~Pe\~na,
  Phys.\ Rev.\ D {\bf 101}, 114008 (2020)
  [arXiv:2003.04850 [hep-ph]].





\bibitem{Przygoda16} 
  W.~Przygoda [HADES Collaboration],
  JPS Conf.\ Proc.\  {\bf 10}, 010013 (2016).



\bibitem{Scozzi17} 
  F.~Scozzi [HADES Collaboration],
  EPJ Web Conf.\  {\bf 137}, 05023 (2017).





\bibitem{Briscoe15} 
  W.~J.~Briscoe, M.~Doring, H.~Haberzettl, D.~M.~Manley, M.~Naruki, 
I.~I.~Strakovsky and E.~S.~Swanson,
  Eur.\ Phys.\ J.\ A {\bf 51}, 129 (2015)
  [arXiv:1503.07763 [hep-ph]].








\bibitem{Pacetti15a} 
  S.~Pacetti, R.~Baldini Ferroli and E.~Tomasi-Gustafsson,
  Phys.\ Rept.\  {\bf 550-551}, 1 (2015).


\bibitem{Aubert07a} 
  B.~Aubert {\it et al.} [BaBar Collaboration],
  Phys.\ Rev.\ D {\bf 76}, 092006 (2007)
  [arXiv:0709.1988 [hep-ex]].



\bibitem{Dobbs17a} 
  S.~Dobbs, K.~K.~Seth, A.~Tomaradze, T.~Xiao and G.~Bonvicini,
  Phys.\ Rev.\ D {\bf 96}, 092004 (2017)
  [arXiv:1708.09377 [hep-ex]].



\bibitem{Ablikim18a} 
  M.~Ablikim {\it et al.} [BESIII Collaboration],
  Phys.\ Rev.\ D {\bf 97}, 032013 (2018)
  [arXiv:1709.10236 [hep-ex]].



\bibitem{Singh16a} 
  B.~Singh {\it et al.} [PANDA Collaboration],
  Phys.\ Rev.\ D {\bf 95}, 032003 (2017)
  [arXiv:1610.02149 [nucl-ex]].




\bibitem{Omega3}
  G.~Ramalho, in preparation.














\bibitem{Lalik19} 
  R.~Lalik [HADES Collaboration],
  J.\ Phys.\ Conf.\ Ser.\  {\bf 1137}, 012057 (2019).





\bibitem{Rathod20} 
  N.~Rathod, R.~Lalik, A.~Malige, P.~Salabura and J.~Smyrski,
  Acta Phys.\ Polon.\ B {\bf 51}, 239 (2020).


\bibitem{HADES20} 
   HADES Collaboration, in preparation.




\bibitem{Junker20a} 
  O.~Junker, S.~Leupold, E.~Perotti and T.~Vitos,
  Phys.\ Rev.\ C {\bf 101}, 015206 (2020)
  [arXiv:1910.07396 [hep-ph]].
 
\bibitem{Holmberg18} 
  M.~Holmberg and S.~Leupold,
  Eur.\ Phys.\ J.\ A {\bf 54}, 103 (2018)
  [arXiv:1802.05168 [hep-ph]].





\bibitem{Kaxiras85} 
  E.~Kaxiras, E.~J.~Moniz and M.~Soyeur,
  Phys.\ Rev.\ D {\bf 32}, 695 (1985).



\bibitem{Williams93} 
  R.~A.~Williams, C.~R.~Ji and S.~R.~Cotanch,
  Phys.\ Rev.\ C {\bf 48}, 1318 (1993).



\bibitem{Sakurai60}
  J.~J.~Sakurai,
  Annals Phys.\  {\bf 11}, 1 (1960).






\bibitem{Krivoruchenko02}     
  M.~I.~Krivoruchenko, B.~V.~Martemyanov, A.~Faessler and C.~Fuchs,
  Annals Phys.\  {\bf 296}, 299 (2002)
  [arXiv:nucl-th/0110066].



\bibitem{Iachello04} 
  F.~Iachello and Q.~Wan,
  Phys.\ Rev.\ C {\bf 69}, 055204 (2004).





\bibitem{Siegert4} 
  G.~Ramalho,
  Phys.\ Rev.\ D {\bf 100}, 114014 (2019)
  [arXiv:1909.00013 [hep-ph]].




\bibitem{Siegert1} 
  G.~Ramalho,
  Eur.\ Phys.\ J.\ A {\bf 54}, 75 (2018)
  [arXiv:1709.07412 [hep-ph]];
  G.~Ramalho,
  Phys.\ Rev.\ D {\bf 94}, 114001 (2016)
  [arXiv:1606.03042 [hep-ph]].



\bibitem{Siegert2} 
  G.~Ramalho,
  Phys.\ Rev.\ D {\bf 93}, 113012 (2016)
  [arXiv:1602.03832 [hep-ph]].



\bibitem{GlobalFit} 
  G.~Ramalho,
  Eur.\ Phys.\ J.\ A {\bf 55}, 32 (2019)
  [arXiv:1710.10527 [hep-ph]].



\bibitem{Sahoo95} 
  R.~K.~Sahoo, A.~R.~Panda and A.~Nath,
  Phys.\ Rev.\ D {\bf 52}, 4099 (1995).


\bibitem{Wagner98} 
  G.~Wagner, A.~J.~Buchmann and A.~Faessler,
  Phys.\ Rev.\ C {\bf 58}, 1745 (1998)
  [nucl-th/9808005].


\bibitem{Bijker00} 
  R.~Bijker, F.~Iachello and A.~Leviatan,
  Annals Phys.\  {\bf 284}, 89 (2000)
  [nucl-th/0004034].






\bibitem{Keller12}
  D.~Keller {\it et al.}  [CLAS Collaboration],
  Phys.\ Rev.\ D {\bf 85}, 059903 (2012)
  [arXiv:1111.5444 [nucl-ex]].



\bibitem{Darewych83} 
  J.~W.~Darewych, M.~Horbatsch and R.~Koniuk,
  Phys.\ Rev.\ D {\bf 28}, 1125 (1983).









\bibitem{Alepuz16a} 
  H.~Sanchis-Alepuz and C.~S.~Fischer,
  Eur.\ Phys.\ J.\ A {\bf 52}, 34 (2016)
  [arXiv:1512.00833 [hep-ph]].


\bibitem{Alepuz18} 
  H.~Sanchis-Alepuz, R.~Alkofer and C.~S.~Fischer,
  Eur.\ Phys.\ J.\ A {\bf 54}, 41 (2018)
  [arXiv:1707.08463 [hep-ph]].




\bibitem{Leinweber93} 
  D.~B.~Leinweber, T.~Draper and R.~M.~Woloshyn,
  Phys.\ Rev.\ D {\bf 48}, 2230 (1993)
  [hep-lat/9212016].






\bibitem{Aliev06} 
  T.~M.~Aliev and A.~Ozpineci,
  Nucl.\ Phys.\ B {\bf 732}, 291 (2006)
  [hep-ph/0406331].


\bibitem{Wang09}
  L.~Wang and F.~X.~Lee,
  Phys.\ Rev.\  D {\bf 80}, 034003 (2009)
  [arXiv:0905.1944 [hep-ph]].




\bibitem{Schat95} 
  C.~L.~Schat, C.~Gobbi and N.~N.~Scoccola,
  Phys.\ Lett.\ B {\bf 356}, 1 (1995)
  [hep-ph/9506227].


\bibitem{Haberichter97} 
  T.~Haberichter, H.~Reinhardt, N.~N.~Scoccola and H.~Weigel,
  Nucl.\ Phys.\ A {\bf 615}, 291 (1997)
  [hep-ph/9610484].





\bibitem{Kim20a} 
  J.~Y.~Kim and H.~C.~Kim,
  arXiv:2002.05980 [hep-ph].



\bibitem{Butler93a}
  M.~N.~Butler, M.~J.~Savage and R.~P.~Springer,
  Nucl.\ Phys.\ B {\bf 399}, 69 (1993)
  [hep-ph/9211247].


\bibitem{Arndt04} 
  D.~Arndt and B.~C.~Tiburzi,
  Phys.\ Rev.\ D {\bf 69}, 014501 (2004)
  [hep-lat/0309013].






\bibitem{Lebed11}
  R.~F.~Lebed and R.~H.~TerBeek,
  Phys.\ Rev.\ D {\bf 83}, 016009 (2011)
  [arXiv:1011.3237 [hep-ph]].








\bibitem{NDelta} 
  G.~Ramalho, M.~T.~Pe\~na and F.~Gross,
  Eur.\ Phys.\ J.\ A {\bf 36}, 329 (2008)
  [arXiv:0803.3034 [hep-ph]].


\bibitem{NDeltaD} 
  G.~Ramalho, M.~T.~Pe\~na and F.~Gross,
  Phys.\ Rev.\ D {\bf 78}, 114017 (2008)
  [arXiv:0810.4126 [hep-ph]].



\bibitem{LatticeD} 
  G.~Ramalho and M.~T.~Pe\~na,
  Phys.\ Rev.\ D {\bf 80}, 013008 (2009).
  [arXiv:0901.4310 [hep-ph]].




\bibitem{Timelike} 
  G.~Ramalho and M.~T.~Pe\~na,
  Phys.\ Rev.\ D {\bf 85}, 113014 (2012)
  [arXiv:1205.2575 [hep-ph]].




\bibitem{Nucleon} 
  F.~Gross, G.~Ramalho and M.~T.~Pe\~na,
  Phys.\ Rev.\ C {\bf 77}, 015202 (2008)
  [nucl-th/0606029].




\bibitem{Omega} 
  G.~Ramalho, K.~Tsushima and F.~Gross,
  Phys.\ Rev.\ D {\bf 80}, 033004 (2009)
  [arXiv:0907.1060 [hep-ph]].







\bibitem{Nucleon2} 
  F.~Gross, G.~Ramalho and M.~T.~Pe\~na,
  Phys.\ Rev.\ D {\bf 85}, 093005 (2012)
  [arXiv:1201.6336 [hep-ph]].





\bibitem{OctetAxial} 
  G.~Ramalho and K.~Tsushima,
  Phys.\ Rev.\ D {\bf 94}, 014001 (2016)
  [arXiv:1512.01167 [hep-ph]].




 \bibitem{Delta1600} 
  G.~Ramalho and K.~Tsushima,
  Phys.\ Rev.\ D {\bf 82}, 073007 (2010)
  [arXiv:1008.3822 [hep-ph]].




\bibitem{Jido12} 
  G.~Ramalho, D.~Jido and K.~Tsushima,
  Phys.\ Rev.\ D {\bf 85}, 093014 (2012)
  [arXiv:1202.2299 [hep-ph]].



\bibitem{Roper}
  G.~Ramalho and K.~Tsushima,
  Phys.\ Rev.\ D {\bf 81}, 074020 (2010)
  [arXiv:1002.3386 [hep-ph]];
  G.~Ramalho and K.~Tsushima,
  Phys.\ Rev.\ D {\bf 89}, 073010 (2014)
  [arXiv:1402.3234 [hep-ph]].




\bibitem{Nexcitations}
  G.~Ramalho and M.~T.~Pe\~na,
  Phys.\ Rev.\ D {\bf 95}, 014003 (2017)
  [arXiv:1610.08788 [nucl-th]];
  G.~Ramalho and M.~T.~Pe\~na,
  Phys.\ Rev.\ D {\bf 84}, 033007 (2011)
  [arXiv:1105.2223 [hep-ph]].


\bibitem{DeltaFF}
  G.~Ramalho and M.~T.~Pe\~na,
  J.\ Phys.\ G {\bf 36}, 085004 (2009)
  [arXiv:0807.2922 [hep-ph]];
  G.~Ramalho, M.~T.~Pe\~na and F.~Gross,
  Phys.\ Rev.\ D {\bf 81}, 113011 (2010)
  [arXiv:1002.4170 [hep-ph]].
  G.~Ramalho and M.~T.~Pe\~na,
  Phys.\ Rev.\ D {\bf 83}, 054011 (2011)
  [arXiv:1012.2168 [hep-ph]].


\bibitem{SRapp}
  G.~Ramalho,
  Phys.\ Rev.\ D {\bf 95}, 054008 (2017)
  [arXiv:1612.09555 [hep-ph]];
  G.~Ramalho,
  Phys.\ Rev.\ D {\bf 90}, 033010 (2014)
  [arXiv:1407.0649 [hep-ph]].




\bibitem{Lattice} 
  G.~Ramalho and M.~T.~Pe\~na,
  J.\ Phys.\ G {\bf 36}, 115011 (2009)
  [arXiv:0812.0187 [hep-ph]].







\bibitem{Krivoruchenko01}
  M.~I.~Krivoruchenko and A.~Faessler,
  Phys.\ Rev.\  D {\bf 65}, 017502 (2001)
  [arXiv:nucl-th/0104045].




\bibitem{Wolf90}
  G.~Wolf, G.~Batko, W.~Cassing, U.~Mosel, K.~Niita and M.~Schaefer,
  Nucl.\ Phys.\  A {\bf 517}, 615 (1990).






\bibitem{Jones73} 
  H.~F.~Jones and M.~D.~Scadron,
  Annals Phys.\  {\bf 81}, 1 (1973).



\bibitem{Devenish76} 
  R.~C.~E.~Devenish, T.~S.~Eisenschitz and J.~G.~Korner,
  Phys.\ Rev.\ D {\bf 14}, 3063 (1976).







\bibitem{Gross}
  F.~Gross,
  Phys.\ Rev.\  {\bf 186}, 1448 (1969);
  A.~Stadler, F.~Gross and M.~Frank,
  Phys.\ Rev.\  C {\bf 56}, 2396 (1997)
  [arXiv:nucl-th/9703043].







\bibitem{Carlson}
  C.~E.~Carlson and N.~C.~Mukhopadhyay,
  Phys.\ Rev.\ Lett.\  {\bf 81}, 2646 (1998)
  [hep-ph/9804356];
  C.~E.~Carlson,
  Phys.\ Rev.\ D {\bf 34}, 2704 (1986);
    C.~E.~Carlson,
  Few Body Syst.\ Suppl.\  {\bf 11}, 10 (1999)
  [hep-ph/9809595].



\bibitem{Lin09} 
  H.~W.~Lin and K.~Orginos,
  Phys.\ Rev.\ D {\bf 79}, 074507 (2009)
  [arXiv:0812.4456 [hep-lat]].



\bibitem{Boinepalli09} 
  S.~Boinepalli, D.~B.~Leinweber, P.~J.~Moran, A.~G.~Williams, J.~M.~Zanotti and J.~B.~Zhang,
  Phys.\ Rev.\ D {\bf 80}, 054505 (2009)
  [arXiv:0902.4046 [hep-lat]].




\bibitem{JDiaz07} 
  B.~Julia-Diaz, T.-S.~H.~Lee, T.~Sato and L.~C.~Smith,
  Phys.\ Rev.\ C {\bf 75}, 015205 (2007)
  [nucl-th/0611033].











\bibitem{Thomas84} 
  A.~W.~Thomas,
  Adv.\ Nucl.\ Phys.\  {\bf 13}, 1 (1984).


\bibitem{Theberge83} 
  S.~Theberge and A.~W.~Thomas,
  Nucl.\ Phys.\ A {\bf 393}, 252 (1983).

\bibitem{Yamaguchi89} 
  T.~Yamaguchi, K.~Tsushima, Y.~Kohyama and K.~Kubodera,
  Nucl.\ Phys.\ A {\bf 500}, 429 (1989).




\bibitem{Donges95} 
  H.~C.~Donges, M.~Schafer and U.~Mosel,
  Phys.\ Rev.\ C {\bf 51}, 950 (1995)
  [nucl-th/9407012].







\bibitem{Hanhart12} 
  C.~Hanhart,
  Phys.\ Lett.\ B {\bf 715}, 170 (2012)
  [arXiv:1203.6839 [hep-ph]].



\bibitem{Herrmann93} 
  M.~Herrmann, B.~L.~Friman and W.~Norenberg,
  Nucl.\ Phys.\ A {\bf 560}, 411 (1993).
















 

\bibitem{PDG10}
  K.~Nakamura {\it et al.}  [Particle Data Group],
  J.\ Phys.\ G {\bf 37}, 075021 (2010).


\bibitem{Keller11a}
  D.~Keller {\it et al.}  [CLAS Collaboration],
  Phys.\ Rev.\ D {\bf 83}, 072004 (2011)
  [arXiv:1103.5701 [nucl-ex]].





\bibitem{Colas75} 
  J.~Colas, C.~Farwell, A.~Ferrer and J.~Six,
  Nucl.\ Phys.\ B {\bf 91}, 253 (1975).





\bibitem{Molchanov04}
  V.~V.~Molchanov {\it et al.}  [SELEX Collaboration],
  Phys.\ Lett.\ B {\bf 590}, 161 (2004)
  [hep-ex/0402026].




\bibitem{Ablikim19} 
  M.~Ablikim {\it et al.},
  Phys.\ Rev.\ D {\bf 101}, 012004 (2020)
  [arXiv:1911.06669 [hep-ex]].



\bibitem{Muhlich} 
  P.~M\"uhlich and U.~Mosel,
  Nucl.\ Phys.\ A {\bf 773}, 156 (2006)
  [nucl-th/0602054];
  P.~M\"uhlich, Ph.~D.~thesis, University of Giessen, 2007.







\bibitem{Connell97} 
  H.~B.~O'Connell, B.~C.~Pearce, A.~W.~Thomas and A.~G.~Williams,
  Prog.\ Part.\ Nucl.\ Phys.\  {\bf 39}, 201 (1997)
  [hep-ph/9501251].

  
\bibitem{GammaR1} 
  G.~J.~Gounaris and J.~J.~Sakurai,
  Phys.\ Rev.\ Lett.\  {\bf 21}, 244 (1968);
  H.~B.~O'Connell, B.~C.~Pearce, A.~W.~Thomas and A.~G.~Williams,
  Phys.\ Lett.\ B {\bf 354}, 14 (1995)
  [hep-ph/9503332].



\bibitem{GammaR3} 
  M.~Benayoun, M.~Feindt, M.~Girone, A.~Kirk, P.~Leruste, J.~L.~Narjoux and K.~Safarik,
  Z.\ Phys.\ C {\bf 58}, 31 (1993).

\bibitem{GammaR4} 
  D.~M.~Manley and E.~M.~Saleski,
  Phys.\ Rev.\ D {\bf 45}, 4002 (1992).







\end{thebibliography}
\end{document}